\documentclass[11pt, a4paper]{article}

\newcommand{\Blackhat}{{\sc BlackHat}}

\usepackage{jheppub}
\usepackage{amsmath}
\usepackage{amsfonts}
\usepackage{amssymb}
\usepackage{tikz}
\usetikzlibrary{arrows,shapes}

\usepackage[compat=1.1.0]{tikz-feynman}

\usepackage{amssymb}
\usepackage{mathtools}
\usepackage{color}
\usepackage{graphicx}
\usepackage{hhline}
\usepackage{booktabs}

\usepackage{pdfpages}

\usepackage[frozencache, cachedir=./]{minted}  
\makeatletter
\usepackage{etoolbox}
\AtBeginEnvironment{Verbatim}{
   \def\PYGvs@tok@err {\color{black} \def\PYGvs@bc##1{\strut ##1}}
}
\makeatother

\AtBeginEnvironment{minted}{%
  \catcode`?\active
  \begingroup\lccode`~=`\?\lowercase{\endgroup\def~{\linebreak}}%
}

\setmintedinline{fontsize=\footnotesize}

\usepackage{caption}

\usepackage{varwidth}
\DeclareCaptionFormat{mycaptionformat}{%
  \begin{varwidth}{\linewidth}%
    \centering
    #1#2#3%
  \end{varwidth}%
}
\usepackage[toc, page]{appendix}

\usepackage{iftex}

\ifLuaTeX

\usepackage[T1]{fontenc}
\usepackage{unicode-math}
\else

\usepackage[utf8]{inputenc}
\DeclareUnicodeCharacter{03D5}{\ensuremath{\phi}}
\DeclareUnicodeCharacter{03F5}{\ensuremath{\epsilon}}
\DeclareUnicodeCharacter{2202}{\ensuremath{\partial}}
\DeclareUnicodeCharacter{220F}{\ensuremath{\prod}}
\DeclareUnicodeCharacter{03C9}{\ensuremath{\omega}}
\DeclareUnicodeCharacter{2260}{\ensuremath{\neq}}
\DeclareUnicodeCharacter{22C5}{\ensuremath{\cdot}}
\DeclareUnicodeCharacter{27E8}{\ensuremath{\langle}}
\DeclareUnicodeCharacter{27E9}{\ensuremath{\rangle}}
\DeclareUnicodeCharacter{2192}{\ensuremath{\rightarrow}}
\DeclareUnicodeCharacter{00B2}{\ensuremath{{}^2}}
\DeclareUnicodeCharacter{221A}{\ensuremath{\sqrt{}}}
\DeclareUnicodeCharacter{223C}{\ensuremath{\sim}}
\DeclareUnicodeCharacter{226A}{\ensuremath{\ll}}
\DeclareUnicodeCharacter{00D7}{\ensuremath{\times}}

\usepackage{accsupp}
\DeclareUnicodeCharacter{2081}{\ensuremath{_{\BeginAccSupp{method=hex,unicode,ActualText=2081}1\EndAccSupp{}}}}
\DeclareUnicodeCharacter{2082}{\ensuremath{_{\BeginAccSupp{method=hex,unicode,ActualText=2082}2\EndAccSupp{}}}}
\DeclareUnicodeCharacter{2083}{\ensuremath{_{\BeginAccSupp{method=hex,unicode,ActualText=2083}3\EndAccSupp{}}}}
\DeclareUnicodeCharacter{2084}{\ensuremath{_{\BeginAccSupp{method=hex,unicode,ActualText=2084}4\EndAccSupp{}}}}
\DeclareUnicodeCharacter{2085}{\ensuremath{_{\BeginAccSupp{method=hex,unicode,ActualText=2085}5\EndAccSupp{}}}}
\DeclareUnicodeCharacter{2086}{\ensuremath{_{\BeginAccSupp{method=hex,unicode,ActualText=2086}6\EndAccSupp{}}}}

\usepackage{pmboxdraw, newunicodechar, fancyvrb, graphicx}
\newunicodechar{⎡}{\raisebox{1ex}{\makebox[.5ex]{\textSFi}}}
\newunicodechar{⎤}{\raisebox{1ex}{\makebox[.5em]{\textSFiii}}}
\newunicodechar{│}{\makebox[.5em]{\textSFxi}}
\DeclareUnicodeCharacter{23A2}{\makebox[.5em]{\textSFxi}}
\DeclareUnicodeCharacter{23A5}{\makebox[.5em]{\textSFxi}}
\newunicodechar{⎣}{\raisebox{-1ex}{\makebox[.5em]{\textSFii}}}
\newunicodechar{⎦}{\raisebox{-1ex}{\makebox[.5em]{\textSFiv}}}

\fi

\begin{document}

\preprint{IPPP/19/82}

\title{Analytical amplitudes from numerical solutions of the scattering equations} 

\author{Giuseppe De Laurentis}
\emailAdd{giuseppe.de-laurentis@durham.ac.uk}

\affiliation{Institute for Particle Physics Phenomenology, Durham University}

\date{\today}

\abstract{The CHY formalism for massless scattering provides a cohesive framework for the computation of scattering amplitudes in a variety of theories. It is especially compelling because it elucidates existing relations among theories which are seemingly unrelated in a standard Lagrangian formulation. However, it entails operations that are highly non-trivial to perform analytically, most notably solving the scattering equations. We present a new Python package (\href{https://github.com/GDeLaurentis/seampy}{seampy}) to solve the scattering equations and to compute scattering amplitudes. Both operations are done numerically with high-precision floating-point algebra. Elimination theory is used to obtain solutions to the scattering equations for arbitrary kinematics. These solutions are then applied to a variety of CHY integrands to obtain tree amplitudes for the following theories: Yang-Mills, Einstein gravity, biadjoint scalar, Born-Infeld, non-linear sigma model, Galileon, conformal gravity and $(\text{DF})^2$. Finally, we exploit this high-precision numerical implementation to explore the singularity structure of the amplitudes and to reconstruct analytical expressions which make manifest their pole structure. Some of the expressions for conformal gravity and the $(\text{DF})^2$ gauge theory are new to the best of our knowledge.}


\maketitle

\section{Introduction}

The scattering equations (SE) first appeared in the litterature in the context of string theory in the '70s \cite{Fairlie:1972one, Fairlie:1972two, Fairlie:2008dg} and '80s \cite{Gross:1987ar}. They were more recently rediscovered by Cachazo, He and Yuan (CHY) in a series of pioneering papers \cite{Cachazo:2013iaa, Cachazo:2013hca, Cachazo:2013iea} demonstrating that the SE provide a set of algebraic equations that are key to an alternative formulation of scattering amplitudes at tree level in $d$ dimensions. Shortly afterwards, this framework was proven to reproduce the correct results for $ϕ^3$ and Yang-Mills \cite{Dolan:2013isa}, to generalise to loop level \cite{Adamo:2013tsa, Geyer:2015bja}, and to arise naturally from a worldsheet theory called ambitwistor string \cite{Mason:2013sva}.

In this alternative QFT formulation the kinematic information of the scattering process is encoded in a set of variables describing the location of punctures on the Riemann sphere. The locations of the punctures are related to the external momenta by the SE. Tree-level amplitudes are obtained by integrating over the position of the punctures on the Riemann sphere, while removing a redundancy coming from M\"obius transformations, and imposing the solution of the SE. Alternatively, this integral can be recast as a contour integral around the punctures of the Riemann sphere. The rest of the integrand (called the CHY-integrand) depends on the chosen theory and it has the nice feature of making manifest relations that are hidden in a standard Lagrangian formulation. For instance, the CHY-integrands for Yang-Mills, Einstein gravity and biadjoint scalar theory closely match the KLT relations \cite{Cachazo:2013gna, KAWAI19861}.

The main bottleneck for the study of QFTs following this approach is the factorial growth of the number of solutions to the SE. In general, after accounting for M\"obius redundancy, the the CHY formulae are supported on $(n-3)!$ solutions of the SE. More specifically, at three-point there are no free punctures, at four-point the SE have a single rational solution, and at five-point the there two irrational solutions. At six-point there are six irrational solutions which have been shown to be still algebraic in $d=4$ \cite{Weinzierl:2014vwa}. Starting at seven-point in $d=4$ and at six-point for general $d$ dimensions the solutions can not be expressed in terms of radicals. At the same time tree-level amplitudes are rational functions of the external kinematics for any phase space multiplicity. Clearly some non-trivial simplification has to occur.

There exist also formulae specific to $d = 4$ based on the scattering equations refined by MHV degree \cite{Geyer:2014fka, Spradlin:2009qr, Lipstein:2015rxa}. In this case the counting is different and the number of solutions corresponds to the Eulerian numbers.

An intriguing solution found in the literature \cite{Huang:2015yka, Sogaard:2015dba} to this factorial growth problem is to obtain the sum of residues from the integral yielding the amplitude without explicitly finding the position of the poles. This powerful approach makes the rationality of the amplitude manifest even when the punctures are irrational. However, as the analytical complexity grows with the multiplicity of the scattering process, even this approach seems to require some form of numerical or semi-numerical reconstruction in order to achieve an analytical expression for the amplitude.

In this paper we develop a purely numerical approach, followed by an analytical reconstruction with the strategy of Ref.~\cite{DeLaurentis:2019phz}. To perform this reconstruction we need an implementation of the CHY formulae which is both sufficiently stable in singular limits and that yields amplitudes with enough numerical precision. We provide code that satisfies these criteria in a \verb!Python! package which we called \href{https://github.com/GDeLaurentis/seampy}{seampy} (from ``Scattering equations and amplitudes with Python'').

A publicly available package to compute amplitudes within the CHY framework had already been presented in Ref.~\cite{Farrow:2018cqi}. It is based on the scattering equations refined by MHV degree. However, it was not designed to provide amplitudes with the high precision needed by our reconstruction strategy. Furthermore, although the reconstructed analytical expressions we present in section~\ref{sec:analytical-reconstruction} are specific to $d = 4$, our package provides numerical solutions to the SE in general $d$ dimensions.

This article is organised as follows. In section~\ref{sec:theory} we review parts of the CHY formalism, in particular the polynomial form of the scattering equations \cite{Dolan:2014ega}, their solutions by means of elimination theory \cite{Dolan:2015iln, Cardona:2015ouc}, and a variety of CHY-integrands \cite{Cachazo:2014xea, Azevedo:2017lkz}. The algorithm for solving the SE and the CHY-integrands are implemented in the \verb!Python! package provided, which is presented in section~\ref{sec:python-package}. It provides high-precision floating-point solutions to the SE and numerical amplitudes. In section~\ref{sec:analytical-reconstruction} we make use of this technology to explore the singularity structure of amplitudes and reconstruct explicit expressions in the $d=4$ spinor helicity language. Finally, in section~\ref{sec:conclusions-and-outlook} we give our conclusions and outlook.

\section{The CHY formalism}\label{sec:theory}

In this section we briefly review the theory underlining the CHY formalism and in the next section present its implementation in a \verb!Python! library. For a more thorough introduction to the subject, with explicit step by step derivations, please consider Ref.~\cite{DeLaurentis:2016abc} and the references therein.

Let us consider the tree-level scattering of $n$ massless particles in $d$ dimensions. We denote with $A$ the set $\{1,\dots,n\}$, with $k^\mu_a$ ($a \in A$) the $n$ momenta, and with $z_a$ the $n$ special points of the Riemann sphere called punctures. The map from momentum space to the Riemann sphere, as defined in Ref.~\cite{Cachazo:2013iaa}, is the given by

\begin{equation}\label{eq:momentum-map}
  k^\mu_{a}= \frac{1}{2 \pi i} \oint_{|z-z_{a}| = \epsilon} dz \frac{p^\mu(k, z)}{\prod_{b \in A}(z-z_b)} \; ,
\end{equation}

\noindent where $p^\mu(z)$ are $d$ polynomials with coefficients depending on the momenta and the punctures. The contour is taken to encircle the punctures.

From Eq.~(\ref{eq:momentum-map}) it can be shown, as a consistency condition, that the following equations have to be satisfied

\begin{equation}\label{eq:scattering-equations}
  f_a (z, k) \equiv \sum_{b \in A \backslash \{a\}} \frac{k_a \cdot k_b}{z_a - z_b} = 0, \qquad \forall a \in A \; ,
\end{equation}

\noindent these are the so-called \textit{scattering equations}. As previously mentioned, the SE are invariant under M\"obius transformations $\text{SL}(2,\mathbb{C})$, that is under the following mapping

\begin{equation}\label{eq:mobius-mapping}
  z \rightarrow \zeta = \frac{\alpha z + \beta}{\gamma z + \delta} \; .
\end{equation}

Because Eq.~(\ref{eq:mobius-mapping}) has effectively three free complex parameters, we can fix the position of three of the $n$ punctures. A common choice in the literature, which we follow throughout this work, is given by

\begin{equation}\label{eq:mobius-fix-choice}
  z_1 = \infty, \quad z_2 = 1, \quad z_n = 0 \; .
\end{equation}

Scattering amplitudes for $n$ massless particles $A_n$\footnote{Color ordering is assumed for gauge theories.} are then obtained by integrating a CHY-integrand $I_{\scriptscriptstyle CHY}$\footnote{More details on $I_{\scriptscriptstyle CHY}$ are given in section~\ref{sec:integrands-section}. For now it suffices to say that in general it is a function of the punctures z, the momenta k and the polarisations $ϵ$.}  over the solutions of the SE. This can be achieved either with a normal integral over delta functions, or as a contour integral over the SE. As by prescription

\begin{align}
\textit{A}_n \, &= \, i \int \frac{d^nz}{d^3\omega} \;\; I_{\scriptscriptstyle CHY}(z; k; \epsilon) \prod_{a \in A}\,\mkern-8mu^{'}\delta(f_a(z, k)) \label{eq:CHY-integral} \\
                &= \, i \oint_\textit{O} \frac{d^nz}{d^3\omega} \;\; I_{\scriptscriptstyle CHY}(z; k; \epsilon) \prod_{a \in A}\,\mkern-8mu^{'}\frac{1}{f_a(z,k)} \label{eq:CHY-contour-integral} \; ,
\end{align}

\noindent where the M\"obius measure $d\omega$ and the modified product symbol $\prod\,\mkern-8mu'$ are defined as

\begin{equation}\label{eq:mobius-measure}
  d^3\omega = \frac{dz_rdz_sdz_t}{(z_r-z_s)(z_s-z_t)(z_t-z_s)} \; ,
\end{equation}

\begin{equation}\label{eq:mobius-product}
  \prod_{a \in A}\,\mkern-8mu^{'} = (z_i-z_j)(z_j-z_k)(z_k-z_i)\prod_{a \in A \backslash \{i,j,k\}} \; .
\end{equation}

By substituting Eq.~(\ref{eq:mobius-measure}) and Eq.~(\ref{eq:mobius-product}) back into Eq.~(\ref{eq:CHY-integral}) or Eq.~(\ref{eq:CHY-contour-integral}) it can be shown that the amplitude $A_n$ is invariant under M\"obius transformations. Note that in principle the sets $\{i, j , k\}$ and $\{r, s, t\}$ are independent, but in practice they are often taken to be the same for convenience sake.  Clearly the requirement of M\"obius invariance also imposes a restriction on the valid CHY-integrands $I_{\scriptscriptstyle CHY}$, as we will see shortly.

We would like to use a purely algebraic approach, as it is more amenable to be implementation as computer code. To achieve this we can recast Eq.~(\ref{eq:CHY-integral}) from an integral to a summation by changing variables from the punctures $z_a$ to the scattering equations $f_a$. This introduces a Jacobian factor, i.e.~the determinant of the Jacobian matrix defined as

\begin{equation}\label{eq:jacobian-matrix}
  ϕ_{ab}\, = \frac{∂ f_a}{∂ z_b} =
  \begin{cases}
    \frac{2 k_a \cdot k_b}{(z_a - z_b)^2} & a \neq b \; ,\\
    - \sum\limits_{j \in A \backslash \{a\}} \frac{2 k_a \cdot k_j}{(z_a - z_j)^2} & a = b \; .
  \end{cases}
\end{equation}

Again, in the spirit of preserving M\"obius invariance, since we have removed punctures $i$, $j$, and $k$ from the above $\delta$-function, we also have to remove the corresponding rows form the Jacobian. Similarly, we are not integrating over $r$, $s$ and $t$, and therefore those columns have to be removed as well. The matrix of Eq.~(\ref{eq:jacobian-matrix}) with rows $i$, $j$, $k$ and columns $r$, $s$, $t$ removed is denoted by $ϕ_{rst}^{ijk}$. In the end, the relevant Jacobian for the change of variables, which is independent of the M\"obius fixing choice, is given by\footnote{We are also including in the Jacobian $J$ the products of differences of punctures from Eq.~(\ref{eq:mobius-measure}) and Eq.~(\ref{eq:mobius-product}).}

\begin{equation}\label{eq:jacobian}
  \mathit{J} = \frac{(z_i-z_j)(z_j-z_k)(z_k-z_i)(z_r-z_s)(z_s-z_t)(z_t-z_r)}{\det(ϕ_{rst}^{ijk})} \; .
\end{equation}

\noindent If we impose the choice made in Eq.~(\ref{eq:mobius-fix-choice}), we have

\begin{equation}\label{eq:mobius-choice}
  \{i, j, k\} = \{r, s, t\} = \{z_1, z_2, z_n\} = \{\infty, 1, 0\} \; .
\end{equation}

\noindent We now write Eq.~(\ref{eq:CHY-integral}) for the scattering amplitudes as

\begin{equation}\label{eq:amplitudes-as-sum}
  \textit{A}_n \, = \, z_1^4 \cdot i \sum_{j = 1}^{(n-3)!} \frac{I_{\scriptscriptstyle CHY}(z^{(j)}(k); k; ϵ)}{\det(ϕ_{rst}^{ijk})(z^{(j)}(k); k)}  \; ,
\end{equation}

\noindent where $j$ labels the solution of to the SE given by the set of punctures $z^{(j)}$, which are themselves function of the momenta $k$. Note that, because of Eq.~(\ref{eq:jacobian}) and our choice Eq.~(\ref{eq:mobius-choice}), the Jacobian $\mathit{J}$ introduces the four powers of $z_1 = \infty$ in the numerator. Therefore, $I_{\scriptscriptstyle CHY}$ must come with four powers of $z_1$ in the denominator for Eq.~(\ref{eq:amplitudes-as-sum}) to be sensible. This is a check of Mobius invariance.

\subsection{Polynomial form of the SE and their solutions}

We now turn to the problem of actually finding the solutions to the SE. It is easiest to consider the SE in the form found in Ref.~\cite{Dolan:2014ega}, where the SE are reformulated as $n - 3$ polynomial equations. We can then follow Ref.~\cite{Dolan:2015iln, Cardona:2015ouc} in using an elimination theory algorithm to find the solutions.

The SE in polynomial form, which are equivalent to the original SE of Eq.~(\ref{eq:scattering-equations}), are given by

\begin{equation}\label{eq:polynomial-se}
  h_{m} = \sum_{S \subset A',\, |S| = m} k^2_{S_1}z_S=0 \, , \:\: \mbox{with} \:\: 1 \leq m \leq n - 3 \, ,
\end{equation}

\noindent where the sets $A'$ and $S_1$ are defined as
\begin{equation}
  A' = A \backslash \{1,n\} \, , \:\: S_1 = S \cup \{1\} 
\end{equation}

\noindent and where $k_S$ and $z_S$ are defined as
\begin{equation}
  \quad k_S = \sum_{b \in S} k_b \:\:\: \mbox{and} \:\:\: z_S = \prod_{b \in S} z_b \; .
\end{equation}

\noindent In the above $z_1$ and $z_n$ have already been set to $\infty$ and $0$ respectively, but $z_2$ is still kept free.

This is a system of $n - 3$ polynomial equations ($h_{1 \leq m \leq n - 3}$) in $n - 2$ variables ($z_{2 \leq i \leq n-1}$). As such it can be solved by using an elimination theory algorithm. The idea underpinning elimination theory is to express the system of equations in matrix form and to introduce more variables and equations until the system is over-specified and yields a consistency condition in the form of $\det(M_n) = 0$. Here we are going to discuss directly the general $n$ case. A more detailed discussion can be found in the original papers of Ref.~\cite{Dolan:2015iln, Cardona:2015ouc} or in Ref.~\cite{DeLaurentis:2016abc}.

In general, the aim is to obtain an equation of order $(n - 3)!$ in the ratio $z_{n - 1} / z_{n - 2}$. The original set of $2^{n-4}$ monomials we wish to eliminate is given by

\begin{equation}
  V^T = \{1, z_2\} \times \{1, z_3\}  \times \, ... \, \times \{1, z_{n-3}\} \; .
\end{equation}

\noindent We introduce a auxiliary set

\begin{equation}
  W^T = \{1\} \times \{1, z_3\} \times \{1, z_4, z_4^2\}  \times \, ... \, \times \{1, z_{n-3} , ... , z_{n-3}^{n-5}\} \; ,
\end{equation}

\noindent which contains $(n-4)!$ terms. The new set of monomials is then given by

\begin{equation}\label{eq:extended-V}
  V^T \rightarrow V^T \times W^T = \{1, z_2\} \times \{1, z_3, z_3^2\} \times \, ... \, \times \{1, z_{n-3} , ... , z_{n-3}^{n-4}\} \; ,
\end{equation}

\noindent which is of length $(n-3)!$. Similarly, the new $(n-3)!$ equations are given by

\begin{equation}\label{eq:extended-H}
  H^T \rightarrow H^T \times W^T \; ,
\end{equation}

\noindent where $H^T$ denotes the vector of polynomial scattering equations $h_{1 \leq m \leq n - 3}$.

This procedure ensures that the number of monomials matches the number of equations, thus allowing to express the system in matrix form.

Then, by taking partial derivatives of the entries of the extended $H$ of Eq.~(\ref{eq:extended-H}) w.r.t.~those of the extended $V$ of Eq.~(\ref{eq:extended-V}), we could construct the $(n-3)!\times(n-3)!$ matrix $M_n$ whose determinant is the required equation. However, this is not necessary in practice since the matrix $M_n$ can be built recursively in a block-matrix format starting directly from the original set $h_{1 \leq m \leq n - 3}$ and their derivatives w.r.t.~$z_{2 \leq i \leq n-3}$. We denote the derivatives with superscripts ($M^z = \partial_z M$) and we have

\begin{equation}\label{eq:matrix-recursion}
  M_i=
  \left(
    \begin{array}{ccccccc}
      M_{i-1} & M_{i-1}^{z_{i-3}} & 0 & \dots & 0 & 0\\
      0 & M_{i-1} & M_{i-1}^{z_{i-3}} & \dots & 0 & 0\\
      \vdots & \vdots & \vdots & \ddots & \vdots & \vdots\\
      0 & 0 & 0 & \dots & M_{i-1} & M_{i-1}^{z_{i-3}}\\
    \end{array}
  \right), \quad
  M_4=H, \quad
  H =
  \left(
    \begin{array}{c}
      h_1\\
      h_2\\
      \vdots\\
      h_{n-3}\\
    \end{array}
  \right) \; ,
\end{equation}

\vspace{2mm}

\noindent with $M_i$ of dimensions $(i-4)\times(i-3)$ when written in terms of $M_{i-1}$. After the derivative is taken $z_{i-3}$ is set to zero. $M_n$ is then a function of $z_{n - 1}$ and $z_{n - 2}$ only, the required equation of order $(n-3)!$ in $z_{n - 1} / z_{n - 2}$ is simply $\det(M_n) = 0$, and its roots are the solutions we seek. Note that, as discussed in the introduction, it is feasible to perform this root-finding step analytically only for low phase space multiplicities.

Clearly we are not at the end of the calculation yet, because we want values or expressions for the punctures themselves not for ratios. This is achieved by reintroducing one variable at a time in $M$. More explicitly, we first check with Eq.~(\ref{eq:extended-V}) the position in the vector of the variable $\tilde z$ we want to reintroduce (say it is the $j^{th}$ entry) then we add $\tilde z$ times the $j^{th}$ column of $M$ to its first column, and eventually remove the $j^{th}$ column and the last row. This leads to a matrix of size $(n-3)! - 1 \times (n-3)! - 1$ whose determinant will be a linear equation for $\tilde z$. There is one notable exception to this procedure, namely when $\tilde z = z_2$ we set $z_2 = 1$ and get a linear equation for $z_{n - 2}$ instead.

Finally, we are left with $(n-3)!$ sets of punctures $\{z_1 = \infty,\, z_2 = 1, \, z_3, \, \dots \, z_{n - 1}, \, z_n = 0\}$ that solve the scattering equations.

\subsection{CHY-integrands}\label{sec:integrands-section}

So far we have treated the theory-independent part of Eq.~(\ref{eq:amplitudes-as-sum}). Now we consider the theory-dependent term $I_{\scriptscriptstyle CHY}$. It can be built in a modular way from various building blocks. Here we review the definition of some of those building blocks found in Ref.~\cite{Cachazo:2014xea} and in Ref.~\cite{Azevedo:2017lkz} which we have implemented in the \verb!Python! package presented in the next section.

Starting from the building blocks that are matrices, we have the $2n \times 2n$ anti-symmetric matrix $\Psi$ which is defined block-wise in terms of two $n \times n$ anti-symmetric matrices $A$ and $B$ and in terms of a third $n \times n$ matrix $C$. The definitions follow.

\begin{equation}\label{eq:Psi-A}
  \Psi =
  \left(
    \begin{array}{cc}
      A & -C^T\\
      C & B
    \end{array}
  \right),
  \quad
  A_{ab} =
  \begin{cases}
    \frac{2 k_a \cdot k_b}{(z_a - z_b)} & a \neq b \; ,\\
    0 & a = b \; ,
  \end{cases}
\end{equation}

\begin{equation}
  B_{ab} =
  \begin{cases}
    \frac{2 ϵ _a \cdot ϵ_b}{(z_a - z_b)} & a \neq b \; ,\\
    0 & a = b \; ,
  \end{cases} 
  \quad
  C_{ab} =
  \begin{cases}
    \frac{2 ϵ_a \cdot k_b}{(z_a - z_b)} & a \neq b \; ,\\
    - \sum\limits_{j \in A \backslash \{a\}} \frac{2 ϵ_a \cdot k_j}{(z_a - z_j)} & a = b \; .
  \end{cases}
\end{equation}

Since these are matrices we have to define an operation which converts them to a rank-one object before we can use them to construct $I_{\scriptscriptstyle CHY}$. In the case of anti-symmetric matrices the determinant can be written as a square of a polynomial in the matrix entries. This polynomial is called the Pfaffian and it was shown to be the correct operation to perform. More specifically, since the matrix $\Psi$ has two null vectors and its Pfaffian would be zero, it is necessary to define a reduced Pfaffian $\text{PF}'$ as

\begin{equation}
  \text{PF}'(\Psi) = \frac{(-1)^{i+j}}{z_i - z_j} \text{PF}(\Psi_{ij}^{ij}) \; ,
\end{equation}

\noindent where $\Psi_{ij}^{ij}$ again denotes deletion of rows and columns $i$ and $j$. The same reduction applies also to different arguments, such as the matrix $A$.

We also consider two scalar building blocks $C_n$ and $W_1$.  $C_n$ is a cyclic Parke-Taylor-like factor simply defined as

\begin{equation}\label{eq:cyclic-factor}
  C_n = \frac{1}{(z_1 - z_2) \dots (z_n - z_1)} \; ,
\end{equation}

\noindent and the $W_1$ function is defined as \footnote{We use $W_1$ to denote the function $W_{1...1}$ from Ref.~\cite{Azevedo:2017lkz}.}

\begin{equation}
  W_1 = ∏_{i \in A} ω_i \;, \quad \text{with} \quad ω_i = \sum\limits_{j \in A \backslash \{i\}} \frac{ϵ_i \cdot k_j \, (z_j - z_r)}{(z_r - z_i)(z_i - z_j)} \;, \;\; r ≠ i.
\end{equation}

$I_{\scriptscriptstyle CHY}$ is built from products of pairs of these building blocks. A more detailed analysis reveals that $\text{PF}'(\Psi)$, $C_n$ and $W_1$ come with a factor of $z_1^{-2}$, while $\text{PF}'(A)$ comes with a factor of $z_1^{-1}$. This dictates which combinations are allowed by M\"obius invariance (recall that overall we need four powers of $z_1$ to balance out those in Eq.~(\ref{eq:amplitudes-as-sum})).

Table~\ref{table:1} summarises the theories that can be built out of $\text{PF}'(\Psi)$, $C_n$, $\text{PF}'(A)^2$ and $W_1$: EG stands for Einstein Gravity, YM for Yang-Mills, BS for Biadjoint Scalar, BI for Born-Infeld, NLSM for Non Linear Sigma Model and CG for Conformal Gravity. The theories labelled with a question mark do not seem to have an agreed upon name, but they are discussed in the reference from which the $W_1$ function is taken.

\begin{table}
  \centering
  \begin{tabular}{|c||c|c|c|c|} 
    \hline
    $\times$ & $\text{PF}'(\Psi)$ & $C_n$ & $\text{PF}'(A)^2$ & $W_1$ \\
    \hhline{|=#=|=|=|=|} 
    $\text{PF}'(\Psi)$ & EG & YM & BI & CG \\
    \hline
    $C_n$ & YM  & BS & NLSM & $(\text{DF})^2$\\
    \hline
    $\text{PF}'(A)^2$ & BI & NLSM & Galileon & ? \\
    \hline
    $W_1$ & CG & $(\text{DF})^2$ & ? & ? \\
    \hline
  \end{tabular}
  \caption{Possible QFTs built out of $\text{PF}'(\Psi)$ ,  $C_n$ ,  $\text{PF}'(A)^2$  and  $W_1$. \newline A product is implied between rows and columns, eg: $I_{\scriptscriptstyle CHY,\; EG} = \text{PF}'(\Psi) \times \text{PF}'(\Psi)$.}
  \label{table:1}
\end{table}

This is by no means a complete recount of all possible integrands $I_{\scriptscriptstyle CHY}$, but it is sufficient to illustrate the framework. Also note that, as anticipated in the introduction, relations among theories through double copies are now manifest in the structure of the integrands.

Finally, remember that the CHY-integrands are not unique. For instance, a different integrand for conformal gravity is given in Ref.~\cite{Farrow:2018yqf}.

\section{Python libraries}\label{sec:python-package}

In this section we introduce two new packages developed in \verb!Phyton 2.7!:
\begin{itemize}
\item[$\circ$] \verb!seampy! (Scattering equations and amplitudes with \verb!Python!),
\item[$\circ$] \verb!lips! ($d=4$ Lortenz invariant phase space).
\end{itemize}
The former provides high-precision floating-point solutions to the scattering equations in $d$ dimensions and a variety of numerical scattering amplitudes built from their solutions. The latter is used to manipulate and pass a high-precision phase space point as input to the numerical amplitude.

Both packages are available on the \href{https://pypi.org/}{Python Package Index}. The source code is available on \href{https://github.com/GDeLaurentis?tab=repositories}{github} and the documentation on the associated github pages \href{https://gdelaurentis.github.io/seampy/}{seampy} and \href{https://gdelaurentis.github.io/lips/}{lips}. Their installation is straightforward thanks to pip:
\vspace{-3mm}
\begin{minted}[escapeinside=||, mathescape=true, linenos=False, numbersep=5pt, gobble=2, frame=lines, framesep=2mm, breaklines, breakautoindent=false, breakindent=-12.5pt, ]{shell}
  pip install --upgrade seampy    # this installs lips as well
  pip install --upgrade lips      # but it can be installed separately
\end{minted}

The same commands can be used to update the libraries. The \verb!--upgrade! option ensures that the lastest version is always used. A review of the key features of these packages is now provided. Further examples are given in section~\ref{sec:analytical-reconstruction} and in the appendices~\ref{sec:appendix1} and~\ref{sec:appendix2}.

\subsubsection{Solving the scattering equations}


In this section we show how to easily obtain solutions for the scattering equations. All the following examples have $n=6$.

The SE in polynomial form as in Eq.~(\ref{eq:polynomial-se}) can be accessed as follows:
\vspace{-3mm}
\begin{minted}[escapeinside=||, mathescape=true, linenos=False, numbersep=5pt, gobble=2, frame=topline, framesep=2mm, breaklines, breakautoindent=false, breakindent=-12.5pt, ]{python}
  >>> hms(6)
\end{minted}
\vspace{-7mm}
\begin{center}
  \noindent\includegraphics{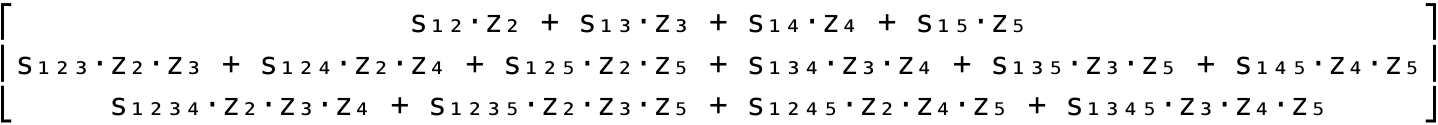}
\end{center}
\vspace{-14mm}
\begin{minted}[escapeinside=||, mathescape=true, linenos=False, numbersep=5pt, gobble=2, frame=bottomline, framesep=2mm, breaklines, breakautoindent=false, breakindent=-12.5pt,
  fontsize=\footnotesize]{python}
\end{minted}

They are functions of the punctures and of Mandelstam invariants, which are given here as they appear in the SE:
\vspace{-3mm}
\begin{minted}[escapeinside=||, mathescape=true, linenos=False, numbersep=5pt, gobble=2, frame=lines, framesep=2mm, breaklines, breakautoindent=false, breakindent=-12.5pt, ]{python}
  >>> punctures(6)
  (z₁, z₂, z₃, z₄, z₅, z₆)
  >>> mandelstams(6)
  (s₁₂, s₁₃, s₁₄, s₁₅, s₁₂₃, s₁₂₄, s₁₂₅, s₁₃₄, s₁₃₅, s₁₄₅, s₁₂₃₄, …)
\end{minted}

The SE can be solved by calling the function \verb!solve_scattering_equations!. It requires two inputs: the multiplicity of the phase space, \verb!n!, and a \verb!Python! dictionary with the numerical values for the Mandelstam invariants, \verb!num_ss!. We therefore need a phase space point. This is easily done through the \href{https://gdelaurentis.github.io/lips/}{lips} toolkit object \verb!Particles! which generates a random phase space point:
\vspace{-3mm}
\begin{minted}[escapeinside=||, mathescape=true, linenos=False, numbersep=5pt, gobble=2, frame=lines, framesep=2mm, breaklines, breakautoindent=false, breakindent=-12.5pt, ]{python}
  >>> oPs = Particles(6)  # arg. is multiplicity of phase space
  >>> num_ss = {str(s): oPs.compute(str(s)) for s in mandelstams(6)}
\end{minted}

Alternatively it is possible to set the momenta from a list by modifying the \verb!four_mom! attribute of each \verb!Particle! in the list subclass \verb!Particles! or to provide an independently construced set of Mandelstam invariants. More of this in appendix \ref{sec:appendix1}.

We can then solve the scattering equations by calling:
\vspace{-3mm}
\begin{minted}[escapeinside=||, mathescape=true, linenos=False, numbersep=5pt, gobble=2, frame=lines, framesep=2mm, breaklines, breakautoindent=false, breakindent=-12.5pt, ]{python}
  >>> sols = solve_scattering_equations(6, num_ss)
\end{minted}

\noindent the output, \verb!sols!, is a list of length $(n-3)!$, in this case 6. Each solution in the list is a dictionary for the non arbitrarily fixed punctures, in this case of the form:
\vspace{-3mm}
\begin{minted}[escapeinside=||, mathescape=true, linenos=False, numbersep=5pt, gobble=2, frame=lines, framesep=2mm, breaklines, breakautoindent=false, breakindent=-12.5pt, ]{python}
  >>> sols[0]
  {'z3': mpc(real='#nbr', imag='#nbr'), ? 'z4': mpc(real='#nbr', imag='#nbr'), ? 'z5': mpc(real='#nbr', imag='#nbr')}
\end{minted}

\noindent where each \kern-.6em\mintinline{python}{'#nbr'} has by default 300 digits of precision.

\subsubsection{Computing scattering amplitudes}

First of all we can list the theories directly available for computation:
\vspace{-3mm}
\begin{minted}[escapeinside=||, mathescape=true, linenos=False, numbersep=5pt, gobble=2, frame=lines, framesep=2mm, breaklines, breakautoindent=false, breakindent=-12.5pt, ]{python}
  >>> theories
  [YM, EG, BS, BI, NLSM, Galileon, CG, DF2]
\end{minted}

To calculate an amplitude we need to generate a phase space point, as in the example for the solutions of the scattering equations:
\vspace{-3mm}
\begin{minted}[escapeinside=||, mathescape=true, linenos=False, numbersep=5pt, gobble=2, frame=lines, framesep=2mm, breaklines, breakautoindent=false, breakindent=-12.5pt, ]{python}
   >>> oParticles = Particles(6)  # arg. is multiplicity of phase space
\end{minted}

We then need to declare what quantity we want to compute. This requires us to specify a theory and a multiplicity. For example, biadjoint scalar theory (BS) amplitudes or non-linear sigma model (NLSM) amplitudes can be accessed as follows:
\vspace{-3mm}
\begin{minted}[escapeinside=||, mathescape=true, linenos=False, numbersep=5pt, gobble=2, frame=lines, framesep=2mm, breaklines, breakautoindent=false, breakindent=-12.5pt, ]{python}
  >>> oBSAmp = NumericalAmplitude(theory='BS', multiplicity=6)
  >>> oNLSMAmp = NumericalAmplitude(theory='NLSM', multiplicity=6)
\end{minted}

Gauge and gravity theories also require an helicity configuration to be specified (the multiplicity is then deduced from it). Note that for gravity theories we are suppressing the repeated helicity sign since we don't have mixed cases such as dilatons. This means that in the following code snippet for conformal gravity (CG) \verb!helconf=pmpmpm! stands for $1^{++}2^{--}3^{++}4^{--}5^{++}6^{--}$.
\vspace{-3mm}
\begin{minted}[escapeinside=||, mathescape=true, linenos=False, numbersep=5pt, gobble=2, frame=lines, framesep=2mm, breaklines, breakautoindent=false, breakindent=-12.5pt, ]{python}
   >>> oDFAmp = NumericalAmplitude(theory='DF2', helconf='pmpmpm')
   >>> oCGAmp = NumericalAmplitude(theory='CG', helconf='pmpmpm')
\end{minted}

It is then simply a matter of evaluating any amplitude at the phase space point:
\vspace{-3mm}
\begin{minted}[escapeinside=||, mathescape=true, linenos=False, numbersep=5pt, gobble=2, frame=lines, framesep=2mm, breaklines, breakautoindent=false, breakindent=-12.5pt, ]{python}
  >>> oBSAmp(oParticles)
  mpc(real='#nbr', imag='#nbr')
\end{minted}

Since most of these helicity amplitudes come with pre-factors of $\sqrt{2}$, we decided to normalise them in such a way that numerical coefficients in analytical expressions are rational fractions and often simply the imaginary unit. This also allows for easier comparison to other codes, which usually adopt such a normalisations. For instance, in the case of Yang-Mills amplitudes the right hand side of Eq.~(\ref{eq:amplitudes-as-sum}) is multiplied by $1/(\sqrt{2})^{n-2}$, so that the numerical coefficient in the Parke-Taylor expression for MHV amplitudes is $i$ instead of $(\sqrt{2})^{n-2}i$, where $n$ is the multiplicity of the process.

\subsubsection{Validations}

A first validation of the code is to check the solutions of the scattering questions. This is simply a matter of inserting each of the solutions back in the polynomial SE and check they vanish to working precision. This can easily be done in practice:
\vspace{-3mm}
\begin{minted}[escapeinside=||, mathescape=true, linenos=False, numbersep=5pt, gobble=2, frame=lines, framesep=2mm, breaklines, breakautoindent=false, breakindent=-12.5pt, ]{python}
  >>> sol = solve_scattering_equations(n, num_ss)[0]
  >>> simplify(hms(n).subs(sol).subs(num_ss).subs({punctures(n)[1]: 1})
  [~10 ** -290, ~10 ** -290, ~10 ** -290]  # for n = 6 there are 3 SE
\end{minted}

Additional checks that don't require independent implementations of amplitudes include checking the little group scalings, mass dimensions, pole structure (more of this in section~\ref{sec:analytical-reconstruction}) or properties such as color ordering. For instance, as a sanity check, we can see that $(\text{DF})^2$ is color ordered whereas conformal gravity is not. This is shown in the following snippet (we are still using the \verb!helconf=pmpmpm! amplitudes declared above):
\vspace{-3mm}
\begin{minted}[escapeinside=||, mathescape=true, linenos=False, numbersep=5pt, gobble=2, frame=lines, framesep=2mm, breaklines, breakautoindent=false, breakindent=-12.5pt, ]{python}
  >>> oNewParticles = oParticles.image("321456")  # swap momenta 1 & 3
  >>> abs(oCGAmp(oParticles) - oCGAmp(oNewParticles)) < 10 ** -270
  True
  >>> abs(oDFAmp(oParticles) - oDFAmp(oNewParticles)) < 10 ** -270
  False
\end{minted}

\noindent However, picking the correct cyclic permutation of the external legs leaves the $(\text{DF})^2$ amplitude unchanged as well.

Finally the most stringent tests come from comparing to independent libraries. We have checked all pure gluon (Yang-Mills) tree amplitudes at 3, 4, 5, 6, and 7 point against \Blackhat~\cite{Berger:2008ag} and Yang-Mills, Einstein and conformal gravity against the code of Ref.~\cite{Farrow:2018cqi}. They all match, that is their ratio differs at most by a normalisation factor fixed by convention. 

\section{Analytical reconstruction}\label{sec:analytical-reconstruction}

We now consider how to recover analytical expressions for the tree-level scattering amplitudes discussed so far. There are several reasons why analytical expressions are preferable to numerical ones, such as execution speed, numerical stability and general understanding of their analytical structure. The same reconstruction technique can be applied to all the theories from Table~\ref{table:1}. In the accompanying files we provide sample analytical amplitudes for all these theories up to six point. The the results are given both in human readable format and as expressions readable by the \verb!S@M! \verb!Mathematica!~package~\cite{Maitre:2007jq}.

In this section, we are going to explicitly discuss only the reconstruction of $(\text{DF})^2$ and conformal gravity amplitudes, since they are the ones with a less well known analytical structure and therefore the most interesting to analyse. These theories are related by a double copy relation, similar to that between Yang-Mills and Einstein gravity, namely: $(\text{DF})^2\times \text{YM} ∼ CG$. $(\text{DF})^2$ and conformal gravity present issues with renormalisability and unitarity, since for instance $(\text{DF})^2$ is built out of dimension-six operators, as implied by the name. Despite this, they are of interest for a few reasons. Namely, one type of conformal gravity arises in Berkovits-Witten twistor string \cite{Berkovits:2004jj}, it is the zero-mass limit of a mass-deformed theory that reproduces Einstein gravity in the infinite-mass limit \cite{Johansson:2018ues}, and it may be useful for computing Einstein gravity amplitude in curved backgrounds for cosmological applications \cite{Maldacena:2011mk, Anastasiou:2016jix}.

More specifically, in the following paragraphs we are going to provide: a) the first complete set of five-point $(\text{DF})^2$ amplitudes (one of which we could confirm numerically with that found in Ref.~\cite{Johansson:2017srf}); b) an alternative expression to that of Ref.~\cite{Berkovits:2004jj} for the five-point MHV conformal gravity amplitude; c) results for the leading three-particle sigularities of the six-point amplitudes in the MHV and NMHV helicity sectors. All the amplitudes we present are written in the spinor helicity language and are free from spurious singularities, unless explicitly stated. We think that, in order to obtain similar complete results at six point, it could be necessary to use spurious sigularities, which introduces a further complication in the analysis.

We make use of the high floating-point precision provided by \href{https://gdelaurentis.github.io/seampy/}{seampy} and follow the strategy introduced in Ref.~\cite{DeLaurentis:2019phz}. Briefly summarised, we study the behaviour of amplitudes in singular limits of complex phase space to obtain the poles and their degree. We then study the amplitudes in doubly singular regions to obtain information about the structure of the denominators of the amplitude. Using this information we generate ans\"atze for the residues of different poles and solve linear systems for the coefficients of bases of spinor expressions in the numerators. If a reconstructed ansatz is correct, once subtracted from the numerical amplitude, it removes a singularity. We repeat the procedure until the amplitude is fully reconstructed.

Explicit examples are discussed in the following subsections.

\subsection{Five-point amplitudes}

\subsubsection{$(\text{DF})^2$: five-point all-plus (explained example)}

In contrast to QCD amplitudes, five-point $(\text{DF})^2$ amplitudes are non zero for all helicity configurations even at tree level. They are color ordered, like QCD, because their CHY-integrand contains the Parke-Taylor-like cyclic factor $C_n$ of Eq.~(\ref{eq:cyclic-factor}). Therefore, the symmetry group is restricted to cyclic and anti-cylic permutations. It can be generated from two operations, which can be thought of as the rotations and reflections of a pentagon (i.e.~the dihedral group $D_5$):

\begin{equation}\label{eq:all-plus-symmetry-group-generators}
  (12345 → 23451) \quad \text{and} \quad (12345 → -15432) \;.
\end{equation}

The minus sign in the reflection comes from the partity operation applied to vector particles ($J^P=1^-)$. In total the group contains 10 elements (including the identity). 

The poles and their order, as well as any common factor in the numerator, can be obtained by studying the behaviour of the amplitude in singular limits. A singular limit is intended as a region of phase space where a single spinor helicity invariant vanishes ($∼O(ϵ ≪ 1)$). We can see how this procedure works in practice in the case of angle and square spinor brackets with the following snippet, which can be run with the provided packages:
\vspace{-3mm}
\begin{minted}[escapeinside=||, mathescape=true, linenos=False, numbersep=5pt, gobble=2, frame=lines, framesep=2mm, breaklines, breakautoindent=false, breakindent=-12.5pt, ]{python}
  >>> from __future__ import unicode_literals
  >>> from lips import Particles
  >>> from seampy import NumericalAmplitude
  >>> import mpmath
  
  >>> oDF2Amp = NumericalAmplitude("DF2", helconf="+++++")
  >>> oParticles = Particles(oDF2Amp.multiplicity)
  >>> oParticles.set("⟨1|2⟩", 10 ** -30)
  >>> a = oDF2Amp(oParticles)
  >>> oParticles.set("⟨1|2⟩", 10 ** -31)
  >>> b = oDF2Amp(oParticles)
  >>> round(mpmath.log(abs(b)/abs(a))/mpmath.log(10))
  2.0  # this is the order of the pole ⟨1|2⟩
\end{minted}

What the above code does is to compute the amplitude at two phase space points and to calculate the slope of the line going through the two points in a log-log plot (Amplitude vs.~spinor invariant).

Following this same procedure with the rest of the spinor invariants we obtain a first look at the analytical structure of the all plus amplitude:
\begin{equation}\label{eq:DF2-all-plus-least-common-denominator}
  A_{(\text{DF})^2}(1^+,\,2^+,\,3^+,\,4^+,\,5^+) = \frac{\mathcal{N}}{⟨12⟩²⟨13⟩⟨14⟩⟨15⟩²⟨23⟩²⟨24⟩⟨25⟩⟨34⟩²⟨35⟩⟨45⟩²} \, ,
\end{equation}

\noindent where $\mathcal{N}$ is some numerator structure.

Two comments are now in order. Firstly, note that the adjacent particle singularities are of second order. This reflects the fact that this theory has a quartic propagator instead of the usual quadratic one. Secondly, although in this case it is possible to obtain an expression for the numerator $\mathcal{N}$, it is often not feasible to do so in this single fraction representation, especially with higher point amplitudes; and even when it is possible, the result is complicated and obscures the structure of the amplitude.

In order to obtain a compact representation, we want to write the amplitude as a sum of fractions, each of which should have a simpler denominator structure than the expression above. It is generally convenient to start by considering the double poles, since they make it difficult to numerically access the corresponding simple poles. We study doubly singular limits, that is regions of phase space where pairs of spinor invariants vanish. In practice, this can be done with the same code snippet as above, by replacing the \verb!oParticles.set! function with the \verb!oParticles.set_pair! one. For example, for the pair $⟨12⟩,\,⟨23⟩$ we have:
\vspace{-8mm}
\begin{minted}[escapeinside=||, mathescape=true, linenos=False, numbersep=5pt, gobble=2, frame=lines, framesep=2mm, breaklines, breakautoindent=false, breakindent=-12.5pt, ]{python}
  >>> oParticles.set_pair("⟨1|2⟩", 10 ** -30, "⟨2|3⟩", 10 ** -30)
\end{minted}

\begin{table}
  \centering
  \begin{tabular}{ccccccccccc}
    \toprule
    {}    & ⟨13⟩ & ⟨14⟩ & ⟨15⟩ & ⟨23⟩ & ⟨24⟩ & ⟨25⟩ & ⟨34⟩ & ⟨35⟩ & ⟨45⟩ \\
    \midrule
    ⟨12⟩ & 2 & 2 & 2 & 2 & 2 & 2 & 3 & 2 & 3 \\
    \bottomrule
  \end{tabular}
  \caption{Doubly singular limits for $⟨12⟩$ in $A_{(\text{DF})^2}(1^+,\,2^+,\,3^+,\,4^+,\,5^+)$}
  \label{table:doubly-singular-limits}
\end{table}

By repeating the same procedure with all pairs involving $⟨12⟩$ and recording the behaviour of the amplitude in the corresponding doubly singular limit we can generate Table~\ref{table:doubly-singular-limits}. Since $⟨12⟩$ is already a double pole, it is not likely for any other invariant appearing with a $2$ in the table to be in the same denominator as $⟨12⟩^2$. Therefore, we make an ansatz where only $⟨34⟩$ and $⟨45⟩$ (as simple poles) appear together with $⟨12⟩^2$. More rigorously, we conjecture that:
\begin{equation}\label{eq:DF2-all-plus-12-limit}
  \lim_{⟨12⟩ → 0} A_{(\text{DF})^2}(1^+,\,2^+,\,3^+,\,4^+,\,5^+) = \frac{\mathcal{N}_{12}}{⟨12⟩²⟨34⟩⟨45⟩} + O(⟨12⟩^{-1}) \, .
\end{equation}
\newpage
To check whether the above is true or not, we start by noting that the amplitude has mass dimension\footnote{Natural units are assumed. In practice the mass dimension can be numerically obtained by re-scaling all the momenta.} of $1$ and little group weights\footnote{Little group transformations modify spinors, while leaving four momenta unchanged. Thus little group scalings can be numerically obtained by re-scaling the spinors. For more details see Ref.~\cite{Elvang:2013cua}.} of $[-2,\, -2,\, -2,\, -2,\, -2]$. Therefore, the numerator in the RHS must have mass dimension $5$ and little group weights $[0,\, 0,\, -1,\, 0,\, -1]$ in order to match the LHS. We then generate a complete set of linearly independent products of spinor invariants consistent with these constraints. In this specific case the basis contains 20 independent entries:
\begin{gather*}
  ⟨12⟩⟨13⟩[13][13][25], \quad ⟨12⟩⟨15⟩[13][15][25], \quad ⟨12⟩⟨23⟩[13][23][25], \quad ⟨12⟩⟨25⟩[13][25][25], \\
  ⟨12⟩⟨35⟩[13][25][35], \quad ⟨13⟩⟨13⟩[13][13][35], \quad ⟨13⟩⟨15⟩[13][15][35], \quad ⟨13⟩⟨23⟩[13][23][35], \\
  ⟨13⟩⟨25⟩[12][35][35], \quad ⟨13⟩⟨25⟩[13][25][35], \quad ⟨13⟩⟨35⟩[13][35][35], \quad ⟨15⟩⟨15⟩[15][15][35], \\
  ⟨15⟩⟨25⟩[15][25][35], \quad ⟨15⟩⟨35⟩[15][35][35], \quad ⟨23⟩⟨23⟩[23][23][35], \quad ⟨23⟩⟨25⟩[23][25][35], \\
  ⟨23⟩⟨35⟩[23][35][35], \quad ⟨25⟩⟨25⟩[25][25][35], \quad ⟨25⟩⟨35⟩[25][35][35], \quad ⟨35⟩⟨35⟩[35][35][35]. \\
\end{gather*}
Note that, the basis would have 290 entries if we were to generate it for the numerator of Eq.~\ref{eq:DF2-all-plus-least-common-denominator}. Moreover, since we are not working in a generic phase space region but in the limit of small $⟨12⟩$, it turns out that 10 of the 20 basis elements only contribute to the $O(⟨12⟩^{-1})$ part of Eq.~\ref{eq:DF2-all-plus-12-limit} and thus can be ignored. We can now generate 10 random phase space points in the $⟨12⟩ → ϵ \ll 1$ region and solve for the coefficients of the 10 elements. The solution has only one non zero coefficient:
\begin{equation}\label{eq:12doublepolesnumerator}
  \mathcal{N}_{12} = i[12]⟨13⟩⟨25⟩[35]^2
\end{equation}

To obtain the remaining four double poles, we can simply symmetrise the expression for the $⟨12⟩$ double pole by applying the following cyclic permutations:
\begin{equation}
  (12345 → 23451), \quad (12345 → 34512), \quad (12345 → 45123), \quad (12345 → 51234).
\end{equation}

Once an expression for a particular pole has been reconstructed, it can be numerically subtracted from the amplitude and the left over quantity will not contain that particular singularity anymore. Its singular limits can then be studied, ans\"atze made and reconstructions performed until all the poles have been successfully obtained and the amplitude fully reconstruced.

The final result for the all plus $(\text{DF})^2$ amplitude follows. On the left hand side we give the amplitude written using the symmetries discussed above. This is the format used throughout the rest of the article. For the sake of clarity, below we reproduce on the right hand side the same expression with the meaning of the symmetries made explicit.

\begin{minipage}{.45\linewidth}
  \centering
  \begin{gather*}
    A_{(\text{DF})^2}(1^+,\,2^+,\,3^+,\,4^+,\,5^+) =\\
    \frac{i[12]⟨13⟩⟨25⟩[35]^2}{⟨12⟩^2⟨34⟩⟨45⟩}+\frac{i[14][24][35]}{⟨12⟩⟨35⟩}+\\[3.3mm]
    (12345 → 23451)\,+\\[4.5mm]
    (12345 → 34512)\,+\\[4.5mm]
    (12345 → 45123)\,+\\[4.5mm]
    (12345 → 51234)\,+\\[3.8mm]
    \frac{2i[15][23]⟨4|1+2|4]}{⟨12⟩⟨34⟩⟨45⟩}+\\
    \frac{2i[12][45]⟨3|1+5|3]}{⟨15⟩⟨23⟩⟨34⟩}+\\
    \frac{2i[12][15][34]}{⟨23⟩⟨45⟩}\phantom{+}
  \end{gather*}
\end{minipage}%
\begin{minipage}{.45\linewidth}
  \centering
  \begin{gather*}
    A_{(\text{DF})^2}(1^+,\,2^+,\,3^+,\,4^+,\,5^+) =\\
    \frac{i[12]⟨13⟩⟨25⟩[35]^2}{⟨12⟩^2⟨34⟩⟨45⟩}+\frac{i[14][24][35]}{⟨12⟩⟨35⟩}+\\
    \frac{i⟨13⟩[14]^2[23]⟨24⟩}{⟨15⟩⟨23⟩^2⟨45⟩}+\frac{i[14][25][35]}{⟨14⟩⟨23⟩}+\\
    \frac{i⟨24⟩[25]^2[34]⟨35⟩}{⟨12⟩⟨15⟩⟨34⟩^2}+\frac{i[13][14][25]}{⟨25⟩⟨34⟩}+\\
    \frac{i[13]^2⟨14⟩⟨35⟩[45]}{⟨12⟩⟨23⟩⟨45⟩^2}+\frac{i[13][24][25]}{⟨13⟩⟨45⟩}+\\
    \frac{i⟨14⟩[15][24]^2⟨25⟩}{⟨15⟩^2⟨23⟩⟨34⟩}+\frac{i[13][24][35]}{⟨15⟩⟨24⟩}+\\
    \frac{2i[15][23]⟨4|1+2|4]}{⟨12⟩⟨34⟩⟨45⟩}+\\
    \frac{2i[12][45]⟨3|1+5|3]}{⟨15⟩⟨23⟩⟨34⟩}+\\
    \frac{2i[12][15][34]}{⟨23⟩⟨45⟩}\phantom{+}
  \end{gather*}
\end{minipage}

\subsubsection{$(\text{DF})^2$: five-point single-minus}

The single minus amplitude has a single element in its symmetry group besides the identity, namely $(12345 → -43215)$, and is slightly more complicated than the all plus one.\nopagebreak
\begin{gather*}
  A_{(\text{DF})^2}(1^+,\,2^+,\,3^+,\,4^+,\,5^-) =\\
  \frac{i/2[23]⟨25⟩^3⟨34⟩[45]}{⟨12⟩⟨14⟩⟨23⟩^2⟨24⟩} + \frac{[23]⟨35⟩(-i/2[12]⟨13⟩⟨25⟩+i/2⟨15⟩[15]⟨35⟩)}{⟨13⟩⟨14⟩⟨23⟩⟨34⟩} + \\
  (12345 → -43215)\, +\\
  \frac{i[12]⟨14⟩⟨15⟩⟨25⟩⟨35⟩[45]}{⟨12⟩^2⟨13⟩⟨34⟩^2}+ \frac{i⟨35⟩\mathcal{N}}{⟨12⟩⟨15⟩⟨23⟩⟨34⟩⟨45⟩}+\\
  \frac{-i[12]⟨14⟩⟨23⟩[24]⟨25⟩⟨45⟩}{⟨12⟩⟨13⟩⟨24⟩⟨34⟩^2} + \frac{-i[14][24]⟨25⟩⟨45⟩}{⟨13⟩⟨23⟩⟨24⟩}+ \frac{-i[13][14]^2[24]}{[15]⟨23⟩[45]}\phantom{+} \; ,
\end{gather*}

\noindent In the above $\mathcal{N}$ is given by\nopagebreak
\begin{align*}
  \mathcal{N} = &([12][13]⟨15⟩^2⟨25⟩+[13]^2⟨15⟩^2⟨35⟩+[12]⟨15⟩[23]⟨25⟩^2\\
                &+[13]⟨15⟩[23]⟨25⟩⟨35⟩+[23]^2⟨25⟩^2⟨35⟩) \; .
\end{align*}

\subsubsection{$(\text{DF})^2$: five-point MHV (adjacent)}

This MHV amplitude is the only one we could already find in the litterature, specifically in Ref.~\cite{Johansson:2017srf}, where it was written in terms of Mandelstam invariants. The expression we provide is more concise, makes its symmetry explicit and is free from spurious singularities. We have numerically checked that the two expressions agree. The one we found follows.\nopagebreak
\begin{gather*}
  A_{(\text{DF})^2}(1^+,\,2^+,\,3^+,\,4^-,\,5^-) =\\
  \frac{i[12]⟨14⟩^2⟨25⟩^2⟨45⟩}{⟨12⟩^2⟨15⟩⟨23⟩⟨34⟩} + \frac{[13]⟨45⟩(i⟨12⟩[12]+i/2⟨13⟩[13]+i⟨14⟩[14])}{⟨12⟩⟨23⟩[45]} + \\
  \frac{i[13]^2⟨14⟩⟨35⟩}{⟨12⟩⟨23⟩[45]} + \frac{-i[12]⟨14⟩⟨25⟩⟨45⟩^2}{⟨12⟩⟨15⟩⟨23⟩⟨34⟩} + \frac{i[12][13]⟨15⟩⟨34⟩}{⟨13⟩⟨23⟩[45]} + \\
  (12345 → -32154)\,+\\
  \frac{i⟨13⟩[13][15][34]⟨45⟩}{⟨12⟩⟨23⟩[45]^2} + \frac{-i[12][13]^2[23]}{[15][34][45]}\phantom{+} \;.
\end{gather*}

\subsubsection{$(\text{DF})^2$: five-point MHV (non-adjacent)}

The following is the last independent five-point amplitude. All others can be obtained by permutations and/or conjugation of the amplitudes presented here.\nopagebreak
\begin{gather*}
  A_{(\text{DF})^2}(1^+,\,2^+,\,3^-,\,4^+,\,5^-) =\\
  \frac{i[12]⟨15⟩^2⟨23⟩⟨35⟩}{⟨12⟩^2⟨14⟩⟨45⟩} + \frac{i[34]⟨35⟩^3}{⟨12⟩⟨15⟩⟨24⟩} + \frac{i[12]⟨23⟩⟨35⟩^2}{⟨12⟩⟨24⟩⟨34⟩} + \\
  (12345 → -21543)\, + \\
  \frac{i⟨35⟩\mathcal{N}}{⟨12⟩⟨14⟩⟨24⟩}+\\
  \frac{i[14][24]⟨35⟩}{⟨12⟩[35]} + \frac{-i[12][14]^2[24]^2}{[15][23][34][45]}\phantom{+}
\end{gather*}

\noindent In the above $\mathcal{N}$ is given by\nopagebreak
\begin{align*}
  \mathcal{N} = &([12]⟨13⟩⟨25⟩+⟨13⟩[13]⟨35⟩+⟨15⟩[15]⟨35⟩\\
                &+⟨23⟩[23]⟨35⟩+⟨25⟩[25]⟨35⟩+2⟨35⟩^2[35])
\end{align*}

\subsubsection{Conformal gravity: five-point MHV}

An all-multiplicities expression for MHV conformal gravity amplitudes exists thanks to work by Berkovits and Witten \cite{Berkovits:2004jj}. Here we present an expression specific to five point which makes manifest the absence of terms with pairs of double poles.\nopagebreak
\begin{gather*}
  A_{CG}(1^{++},\,2^{++},\,3^{++},\,4^{--},\,5^{--}) =\\
  \frac{-i[12]^2⟨24⟩[34]⟨45⟩^5}{⟨12⟩^2⟨23⟩⟨34⟩⟨35⟩}+\frac{i[12]^2[13]⟨15⟩⟨45⟩^4}{⟨12⟩⟨13⟩⟨23⟩⟨35⟩}+\\
  (12345 → 23145) \, + \, (12345 → 31245) \, +\\
  \frac{-2i[12][13][23]⟨45⟩^4}{⟨12⟩⟨13⟩⟨23⟩}\phantom{+}
\end{gather*}

\subsection{Six-point partial results}

\subsubsection{$(\text{DF})^2$: six-point MHV (adjacent) (partial)}

In order to convey the increse in complexity that a six-point amplitude entails here we present an expression for the three-particle double poles as well as for the simple poles of non-adjacent three-particle singularities in a six-point MHV $(\text{DF})^2$ amplitude.
\begin{gather*}
A_{(\text{DF})^2}(1^+,\,2^+,\,3^+,\,4^+,\,5^-,\,6^-) = \\
\footnotesize\frac{i[13][46]⟨56⟩\mathcal{N}_1}{⟨12⟩⟨23⟩⟨45⟩[56]^2s_{123}^2} + \footnotesize\frac{i[12][34]⟨26⟩⟨35⟩\mathcal{N}_2}{⟨12⟩^2[16]⟨34⟩[45]s_{345}^2}+\\
\frac{-i[14][24][35][36]⟨56⟩}{⟨12⟩[56]^2s_{124}} + \frac{-i[12]⟨15⟩⟨25⟩[34][45]⟨46⟩}{⟨12⟩^2⟨34⟩[56]s_{125}} + \\
\frac{i[14]^2[23]⟨26⟩⟨36⟩[46]}{⟨23⟩^2[45][56]s_{145}} + \frac{-i[14][23]⟨26⟩⟨5|1+4|2]}{⟨14⟩⟨23⟩[56]s_{145}} + \\
(123456 → 432165)\,+\\
\frac{-i[12][14]⟨15⟩[34]⟨46⟩}{⟨12⟩⟨34⟩[56]s_{125}} + \frac{-i[13]^2[24]^2⟨25⟩⟨36⟩}{⟨13⟩[16]⟨24⟩[45]s_{245}} + \\
\frac{\mathcal{N}}{⟨12⟩²⟨13⟩⟨14⟩⟨16⟩[16]⟨23⟩²⟨24⟩⟨34⟩²⟨45⟩[45][56]²s_{123}s_{234}s_{345}}\phantom{+}
\end{gather*}

\noindent Where $\mathcal{N}_1$ and $\mathcal{N}_2$ are given by:\nopagebreak
\begin{align*}
  \mathcal{N}_1 \, = \, (&-2⟨12⟩^2[12]^2⟨24⟩[24]-2⟨12⟩^2[12]^2⟨25⟩[25]-2⟨12⟩^2[12][13][24]⟨34⟩\\
                         &-2⟨12⟩^2[12][13][25]⟨35⟩-⟨12⟩^2[12][14][25]⟨45⟩-2⟨12⟩[12]^2⟨13⟩⟨24⟩[34]\\
                         &-2⟨12⟩[12]^2⟨13⟩⟨25⟩[35]-2⟨12⟩[12]⟨13⟩[13]⟨34⟩[34]-2⟨12⟩[12]⟨13⟩[13]⟨35⟩[35]\\
                         &-⟨12⟩[12]⟨13⟩[14][35]⟨45⟩-⟨12⟩[12]^2⟨14⟩⟨25⟩[45]-⟨12⟩[12][13]⟨14⟩⟨35⟩[45]\\
                         &+⟨12⟩[12]⟨23⟩[24][35]⟨45⟩+⟨12⟩[12][23]⟨24⟩⟨35⟩[45]+⟨12⟩[13][23]⟨34⟩⟨35⟩[45]\\
                         &+[12]⟨13⟩⟨23⟩[34][35]⟨45⟩)\\[2mm]
  \mathcal{N}_2 \, = \, (&+3⟨12⟩[12]⟨13⟩[13][34]-2⟨12⟩⟨13⟩[13]^2[24]-⟨12⟩[13]⟨14⟩[14][24]\\
                         &+⟨12⟩[12]⟨15⟩[15][34]-⟨12⟩[13][14]⟨15⟩[25]-⟨12⟩[13]⟨23⟩[23][24]\\
                         &-⟨12⟩[15][23][24]⟨25⟩+⟨13⟩^2[13]^2[34]-⟨13⟩[13]^2⟨15⟩[45]\\
                         &+⟨13⟩[13]⟨15⟩[15][34]+⟨13⟩[13]⟨23⟩[23][34]-⟨13⟩[13][23]⟨25⟩[45]\\
                         &+⟨13⟩[15][23]⟨25⟩[34]-⟨14⟩^2[14]^2[34]-[13]⟨14⟩[14]⟨15⟩[45]\\
                         &-⟨14⟩[14]⟨15⟩[15][34]-⟨14⟩[14]⟨24⟩[24][34]-[13]⟨14⟩[24]⟨25⟩[45]\\
                         &-⟨14⟩[15][24]⟨25⟩[34]-[13]⟨15⟩^2[15][45]-⟨15⟩[15][23]⟨25⟩[45])
\end{align*}

In the above expression $\mathcal{N}$ would contain several thousand terms. It is therefore crucial to identify appropriate ways to perform a partial fraction decomposition, since smaller denominators would in turn imply smaller numerators and thus easier systems of linear equations to generate and solve. However, further studies will be necessary to check whether such a decomposition requires the introduction of spurious singularities, like for NMHV amplitudes in Yang-Mills, and if so what form these spurious poles would take. 

\subsubsection{Conformal gravity: NMHV (partial)}

To conclude, we present an expression for the three-particle double poles in the six-point NMHV conformal gravity amplitude. To the best of our knowledge this is the first analytical result, albeit a partial one, for NMHV conformal gravity amplitudes.\nopagebreak
\begin{gather*}
  A_{CG}(1^{++},\,2^{++},\,3^{++},\,4^{--},\,5^{--},\,6^{--}) = \\
  \footnotesize\frac{i[23]^4⟨56⟩^4\mathcal{N}_1}{⟨15⟩⟨16⟩⟨23⟩^2[24][34][56]^2s_{234}^2}+\\
  (123456 → 312645) \, + (123456 → 231564) \, + (123456 → 312564) \, +\\
  (123456 → 231645) \, + (123456 → 312456) \, + (123456 → 231456) \, +\\
  (123456 → 123645) \, + (123456 → 123564) \, +\\
  \frac{\mathcal{N}}{\begin{gathered}(⟨12⟩²⟨13⟩²⟨14⟩[14]⟨15⟩[15]⟨16⟩[16]⟨23⟩²⟨24⟩[24]⟨25⟩[25]⟨26⟩[26]⟨34⟩[34]\\×⟨35⟩[35]⟨36⟩[36][45]²[46]²[56]²s_{124}s_{125}s_{134}s_{135}s_{145}s_{234}s_{235}s_{245}s_{345})\end{gathered}}\phantom{+}
\end{gather*}

\noindent In the above $\mathcal{N}_1$ is given by\nopagebreak
{\small
\begin{align*}
  \mathcal{N}_1 \, = \, (&-[12]^2⟨13⟩[15]⟨23⟩⟨24⟩^2[36]+[12]⟨13⟩[13][15]⟨23⟩⟨24⟩^2[26]-[12]⟨13⟩[13][15]⟨23⟩⟨24⟩⟨34⟩[36]\\
                              &+⟨13⟩[13]^2[15]⟨23⟩⟨24⟩[26]⟨34⟩+[12]^2⟨14⟩[15]⟨23⟩^2⟨24⟩[36]-[12][13]⟨14⟩[15]⟨23⟩^2⟨24⟩[26]\\
                              &-[12]⟨14⟩[14][15]⟨23⟩⟨24⟩⟨34⟩[36]+[13]⟨14⟩[14][15]⟨23⟩⟨24⟩[26]⟨34⟩-[12][13]⟨23⟩^2⟨24⟩^2[25][26]\\
                              &-2[12][13]⟨23⟩^2⟨24⟩[25]⟨34⟩[36]-[12][13]⟨23⟩^2⟨34⟩^2[35][36]-[12][14]⟨23⟩⟨24⟩^3[25][26]\\
                              &-[12][13]⟨23⟩⟨24⟩^2[24]⟨34⟩[56]-2[12][14]⟨23⟩⟨24⟩^2[25]⟨34⟩[36]+[13][14]⟨23⟩⟨24⟩^2[25][26]⟨34⟩\\
                              &-[12][14]⟨23⟩⟨24⟩⟨34⟩^2[35][36]-[13]^2⟨23⟩⟨24⟩[24]⟨34⟩^2[56]+2[13][14]⟨23⟩⟨24⟩[25]⟨34⟩^2[36]\\
                              &+[13][14]⟨23⟩⟨34⟩^3[35][36]-[12][14]⟨24⟩^3[24]⟨34⟩[56]+[14]^2⟨24⟩^3[25][26]⟨34⟩\\
                              &-[13][14]⟨24⟩^2[24]⟨34⟩^2[56]+2[14]^2⟨24⟩^2[25]⟨34⟩^2[36]+[14]^2⟨24⟩⟨34⟩^3[35][36])\,.
\end{align*}
}%
Here $\mathcal{N}$ would contain even more terms then in the six-point $(\text{DF})^2$ example. Similar expressions where the symmetries of the poles are made manifest are also possible in Einstein gravity amplitudes, for example the following represent the three-particle simple poles in the six-point NMHV sector.\nopagebreak
\begin{gather*}
  A_{EG}(1^{++},\,2^{++},\,3^{++},\,4^{--},\,5^{--},\,6^{--}) =\\
  \frac{-i[12]^3⟨56⟩^3⟨4|1+2|3]^4}{⟨12⟩⟨14⟩[14]⟨24⟩[24]⟨35⟩[35]⟨36⟩[36][56]s_{124}}+\\
  (123456 → 132456) + (123456 → 123546) + (123456 → 132546) \, +\\
  (123456 → 321456) + (123456 → 123654) + (123456 → 321654) \, +\\
  (123456 → 231546) + (123456 → 132645) + \\
  \frac{\mathcal{N}}{⟨12⟩⟨13⟩⟨14⟩[14]⟨15⟩[15]⟨16⟩[16]⟨23⟩⟨24⟩[24]⟨25⟩[25]⟨26⟩[26]⟨34⟩[34]⟨35⟩[35]⟨36⟩[36][45][46][56]}
  \phantom{+}
\end{gather*}

However, this symmetric approach, which is also free from spurious singularities, makes it highly non trivial to obtain the rest of the amplitude (i.e.~the numerator $\mathcal{N}$). Indeed, the compact expressions that we are aware of come from BCFW recursions and have a quite different structure:
\vspace{-10mm}
\begin{gather*}
  A_{EG}(1^{++},\,2^{++},\,3^{++},\,4^{--},\,5^{--},\,6^{--}) =\\
  \frac{-i[23]^7⟨34⟩⟨56⟩^7[56]}{⟨15⟩⟨16⟩[24][34]⟨1|2+4|3]⟨1|2+3|4]⟨5|1+6|2]⟨6|1+5|2]s_{234}}+\\
  \footnotesize\frac{i[24]⟨4|1+2|3]^7\left(\begin{gathered}-⟨12⟩[12]⟨13⟩[35]⟨45⟩+⟨12⟩[13]⟨14⟩[25]⟨35⟩+⟨12⟩[23]⟨24⟩[25]⟨35⟩-⟨12⟩[24]⟨34⟩[35]⟨45⟩\\-⟨13⟩⟨14⟩[14][35]⟨45⟩+[13]⟨14⟩^2⟨35⟩[45]-⟨14⟩⟨24⟩[25][34]⟨35⟩+⟨14⟩[24]⟨25⟩⟨34⟩[35]\end{gathered}\right)}{⟨12⟩^2⟨24⟩[35][36][56]⟨1|2+4|3]⟨1|2+4|5]⟨1|2+4|6]⟨4|1+2|5]⟨4|1+2|6]s_{124}}+\\
  \footnotesize\frac{i[12]^6⟨14⟩⟨56⟩^7\left(\begin{gathered}-⟨12⟩[12][23]⟨35⟩[45]-[12]⟨13⟩[14]⟨15⟩[35]+[12]⟨14⟩[34]⟨35⟩[45]+⟨14⟩[15][24][34]⟨35⟩\\-[12]⟨15⟩⟨23⟩[24][35]-[14]⟨15⟩[24]⟨34⟩[35]+⟨23⟩[24]^2[35]⟨45⟩-[23]⟨24⟩[24]⟨35⟩[45]\end{gathered}\right)}{[14]⟨35⟩⟨36⟩⟨3|1+4|2]⟨3|1+2|4]⟨5|1+4|2]⟨5|1+2|4]⟨6|1+4|2]⟨6|1+2|4]s_{124}}+\\
  \frac{-i[34]⟨56⟩⟨4|1+3|2]^7}{⟨13⟩⟨14⟩[25][26]⟨34⟩[56]⟨1|2+6|5]⟨1|2+5|6]⟨3|1+4|2]s_{134}}+\\
  (123456 → 123546) + (123456 → 123654) \, +\\
  \footnotesize\frac{i[23]s_{123}^7\left(\begin{gathered}⟨12⟩⟨13⟩[14][25]⟨45⟩-⟨12⟩[12]⟨14⟩⟨35⟩[45]+⟨12⟩⟨23⟩[24][25]⟨45⟩+⟨12⟩[23]⟨34⟩⟨35⟩[45]\\+⟨13⟩^2[14][35]⟨45⟩-⟨13⟩[13]⟨14⟩⟨35⟩[45]+⟨13⟩⟨23⟩[25][34]⟨45⟩-⟨13⟩[23]⟨25⟩⟨34⟩[45]\end{gathered}\right)}{⟨12⟩^2⟨23⟩[45][46][56]⟨1|2+3|4]⟨1|2+3|5]⟨1|2+3|6]⟨3|1+2|4]⟨3|1+2|5]⟨3|1+2|6]}\phantom{+}
\end{gather*}\\[-1mm]
We have reproduced this result already known in the literature by applying our analytical reconstruction strategy to a single BCFW factorisation channel at a time, which is significantly simpler than the full amplitude\footnote{In this case a $⟨21]$ shift was used.}. Compared to the previous partial result, we note that this representation manifestly does not contain two-particle Mandelstam invariants, but introduces many spurious singularities and hides the symmetries which were manifest in the above partial result.


The strategy of studying a factorisation channel at a time could prove fruitful also in the case of conformal gravity and $(\text{DF})^2$ amplitudes, but the quartic propagator introduces a significant complication in the BCFW recursion.

In fact, the usual $\textit{A}_L\textit{A}_R/p^2$ factorisation is broken by the presence of higher order poles in the Laurent expansion in the shift parameter. We attempted to achieve such a factorisation by means of a Taylor expansion of the numerator $\textit{A}_L\textit{A}_R$ around the pole. However, this involves taking a derivative with respect to the shift parameter which in turns requires the amplitudes to be well defined in the neighbourhood of the factorisation point. This seems to be equivalent to the factorisation formula (Eq.~2.18) given in Ref.~\cite{Johansson:2018ues}, where the derivative is implicit in the fact that we have to take the zero mass limit of expressions like $(\textit{A}_L(m^2)-\textit{A}_L(0))/m^2$. This would also explain why our approach fails: the amplitudes we use are well defined only exactly at the factorisation point, where the legs are on-shell and massless.

However, we do have the six-point amplitude through the CHY formula and there is no need to generate it recursively from lower point amplitudes. At the same time, we expect single factorisation channels to have an easier analytical structure than the full amplitude. This suggests to still look at the amplitude via the residue theorem:
\begin{equation}
  \frac{1}{2\pi i} \oint \frac{\hat{A}(z)}{z} dz = \hat{A}(0) + \sum_i \frac{\text{Res}\hat{A}(z)|_{z=z_i}}{z_i}.
\end{equation}


We can then study one term in the sum in the RHS at a time. Note that the simultaneous need to generate singular phase space limits and to numerically extract the residue from a Laurent expansion in some cases requires to increase the working numerical of precision.

As an example, let us consider the same $⟨21]$ shift as before, and more specifically the $(2, 3, 4)_L$, $(1, 5, 6)_R$ channel, which for Einstein gravity yields the first term from the previous expression, i.e.:

{\small
\begin{equation*}
  \frac{\text{Res} \, \hat A_{EG}^{\textit{NMHV}}(z) }{z}\Big|_{z=z_{\tiny(2, 3, 4)_L, (1, 5, 6)_R}} = \frac{i[23]^7⟨34⟩⟨56⟩^7[56]}{⟨15⟩⟨16⟩[24][34]⟨1|2+4|3]⟨1|2+3|4]⟨5|1+6|2]⟨6|1+5|2]s_{234}}.
\end{equation*}
}

The same shift in the same channel in the case of conformal gravity instead yields:

{\small
\begin{equation*}
 \frac{\text{Res} \, \hat A_{CG}^{\textit{NMHV}}(z) }{z}\Big|_{z=z_{(2, 3, 4)_L, (1, 5, 6)_R}} = \frac{\mathcal{N}}{\begin{gathered}(⟨12⟩^2⟨13⟩^2⟨15⟩⟨16⟩[24]⟨34⟩^2[34][46]^2[56]^2⟨1|3+4|2]^2⟨1|2+4|3]\\×⟨1|2+3|4]^3⟨5|1+6|2]⟨6|1+5|2]s_{124}^2s_{125}^2s_{234}^2)\end{gathered}}.
\end{equation*}
}

The numerator $\mathcal{N}$, having mass dimension of 46, is unfortunately still too complicated to be determined. We see that the conformal gravity residue has more poles and poles of higher order compared to Einstein gravity one, as well as some spurious singularities of order higher than one. Furthermore, note that for this shift the contour integral vanishes for Einstein gravity but not for conformal gravity. Therefore, in the latter case we would have to include a boundary term coming from the residue at infinity. Some of the other possible shifts have the advantage of vanishing on the contour, but the structure of the residues remains similarly complicated. Further work will be required to see whether a reasonably compact analytical expression can be obtained for these residues.

\section{Conclusion and outlook}\label{sec:conclusions-and-outlook}

In this article we have briefly reviewed the CHY formalism for massless tree-level scattering, and more specifically the problem of solving the scattering equations and applying their solutions to CHY-integrands.

In order to overcome the analytical complexity of the computation, we have developed a \verb!Python! package (\href{https://gdelaurentis.github.io/seampy/}{seampy}) which allows to numerically solve the scattering equations and to computate tree amplitudes with high floating-point precision for the following theories: Yang-Mills, Einstein gravity, biadjoint scalar, Born-Infeld, non-linear sigma model, Galileon, conformal gravity and $(\text{DF})^2$.

Finally, we have discussed how to recover analytical expression in the spinor helicity language from numerical evaluations. In particular, we have presented the first complete set of five-point $(\text{DF})^2$ amplitudes, a new form for the five-point MHV conformal gravity amplitude and a discussion with partial results for six-point amplitudes in both $(\text{DF})^2$ and conformal gravity.

In the accompanying files we have provided sample analytical amplitudes for all mentioned theories up to six point. The the results are given both in human readable format and as expressions readable by the \verb!S@M! \verb!Mathematica!~package. 

Let us remark the fact that despite not all the solutions to the scattering equations are rational (except at three and four point), and in some cases not even be expressible in terms of radicals (beyond six point), the tree-level amplitudes built from them are purely rational functions. This is made clear by reconstructing explicit rational analytical expressions from numerical evaluations. The expressions we obtain are usually compact, with a clear symmetry structure when available and free from spurious singularities, unless explicitly stated.

We have observed that complexity increases significantly from five-point to six-point amplitudes and given explicit examples. In the previous section we discussed ways to look at simpler building blocks rather than the full amplitude all at once, such as a modified BCFW recursion for the quartic propagators of conformal gravity and $(\text{DF})^2$ or a more naive application of the residue theorem. These approaches seem promising, since they can still be carried out numerically while resulting in simpler structures to which apply the analytical reconstruction. However, more remains to be done to make this feasible in practice for the more complicated theories.

Finally, going forward it might be interesting to use this numerical approach to the CHY formalism together with the analytical reconstruction tools to look at other interesting quantities such as double copy structures, BCJ numerators, amplitudes with mixed particle content, and loop-level amplitudes.

\acknowledgments
I would like to thank Yang-Hui He for introducing me to the subject of the scattering equations; Arthur Lipstein and Joseph Farrow for useful discussion on formal aspects of the work; and Daniel Maitre for technical suggestions as well as for the use of his code to generate spinor helicity ans\"atze. I am funded by an STFC PhD scholarship.

\pagebreak
\bibliography{article}

\providecommand{\href}[2]{#2}\begingroup\raggedright\begin{thebibliography}{10}

\bibitem{Fairlie:1972one}
D.~Fairlie and D.~Roberts, \emph{{Dual Models without Tachyons - a New
  Approach}}, {\emph{unpublished Durham preprint PRINT-72-2440} (1972) }.

\bibitem{Fairlie:1972two}
D.~Roberts, \emph{{Mathematical Structure of Dual Amplitudes}},
  {\emph{\href{http://etheses.dur.ac.uk/8662/1/8662\_5593.PDF}{Durham PhD
  thesis}} (1972) }.

\bibitem{Fairlie:2008dg}
D.~B. Fairlie, \emph{{A Coding of Real Null Four-Momenta into World-Sheet
  Coordinates}}, \href{https://doi.org/10.1155/2009/284689}{\emph{Adv. Math.
  Phys.} {\bfseries 2009} (2009) 284689}
  [\href{https://arxiv.org/abs/0805.2263}{{\ttfamily 0805.2263}}].

\bibitem{Gross:1987ar}
D.~J. Gross and P.~F. Mende, \emph{{String Theory Beyond the Planck Scale}},
  \href{https://doi.org/10.1016/0550-3213(88)90390-2}{\emph{Nucl. Phys.}
  {\bfseries B303} (1988) 407}.

\bibitem{Cachazo:2013iaa}
F.~Cachazo, S.~He and E.~Y. Yuan, \emph{{Scattering in Three Dimensions from
  Rational Maps}}, \href{https://doi.org/10.1007/JHEP10(2013)141}{\emph{JHEP}
  {\bfseries 10} (2013) 141} [\href{https://arxiv.org/abs/1306.2962}{{\ttfamily
  1306.2962}}].

\bibitem{Cachazo:2013hca}
F.~Cachazo, S.~He and E.~Y. Yuan, \emph{{Scattering of Massless Particles in
  Arbitrary Dimensions}},
  \href{https://doi.org/10.1103/PhysRevLett.113.171601}{\emph{Phys. Rev. Lett.}
  {\bfseries 113} (2014) 171601}
  [\href{https://arxiv.org/abs/1307.2199}{{\ttfamily 1307.2199}}].

\bibitem{Cachazo:2013iea}
F.~Cachazo, S.~He and E.~Y. Yuan, \emph{{Scattering of Massless Particles:
  Scalars, Gluons and Gravitons}},
  \href{https://doi.org/10.1007/JHEP07(2014)033}{\emph{JHEP} {\bfseries 07}
  (2014) 033} [\href{https://arxiv.org/abs/1309.0885}{{\ttfamily 1309.0885}}].

\bibitem{Dolan:2013isa}
L.~Dolan and P.~Goddard, \emph{{Proof of the Formula of Cachazo, He and Yuan
  for Yang-Mills Tree Amplitudes in Arbitrary Dimension}},
  \href{https://doi.org/10.1007/JHEP05(2014)010}{\emph{JHEP} {\bfseries 05}
  (2014) 010} [\href{https://arxiv.org/abs/1311.5200}{{\ttfamily 1311.5200}}].

\bibitem{Adamo:2013tsa}
T.~Adamo, E.~Casali and D.~Skinner, \emph{{Ambitwistor strings and the
  scattering equations at one loop}},
  \href{https://doi.org/10.1007/JHEP04(2014)104}{\emph{JHEP} {\bfseries 04}
  (2014) 104} [\href{https://arxiv.org/abs/1312.3828}{{\ttfamily 1312.3828}}].

\bibitem{Geyer:2015bja}
Y.~Geyer, L.~Mason, R.~Monteiro and P.~Tourkine, \emph{{Loop Integrands for
  Scattering Amplitudes from the Riemann Sphere}},
  \href{https://doi.org/10.1103/PhysRevLett.115.121603}{\emph{Phys. Rev. Lett.}
  {\bfseries 115} (2015) 121603}
  [\href{https://arxiv.org/abs/1507.00321}{{\ttfamily 1507.00321}}].

\bibitem{Mason:2013sva}
L.~Mason and D.~Skinner, \emph{{Ambitwistor strings and the scattering
  equations}}, \href{https://doi.org/10.1007/JHEP07(2014)048}{\emph{JHEP}
  {\bfseries 07} (2014) 048} [\href{https://arxiv.org/abs/1311.2564}{{\ttfamily
  1311.2564}}].

\bibitem{Cachazo:2013gna}
F.~Cachazo, S.~He and E.~Y. Yuan, \emph{{Scattering equations and
  Kawai-Lewellen-Tye orthogonality}},
  \href{https://doi.org/10.1103/PhysRevD.90.065001}{\emph{Phys. Rev.}
  {\bfseries D90} (2014) 065001}
  [\href{https://arxiv.org/abs/1306.6575}{{\ttfamily 1306.6575}}].

\bibitem{KAWAI19861}
H.~Kawai, D.~Lewellen and S.-H. Tye, \emph{A relation between tree amplitudes
  of closed and open strings},
  \href{https://doi.org/https://doi.org/10.1016/0550-3213(86)90362-7}{\emph{Nuclear
  Physics B} {\bfseries 269} (1986) 1 }.

\bibitem{Weinzierl:2014vwa}
S.~Weinzierl, \emph{{On the solutions of the scattering equations}},
  \href{https://doi.org/10.1007/JHEP04(2014)092}{\emph{JHEP} {\bfseries 04}
  (2014) 092} [\href{https://arxiv.org/abs/1402.2516}{{\ttfamily 1402.2516}}].

\bibitem{Geyer:2014fka}
Y.~Geyer, A.~E. Lipstein and L.~J. Mason, \emph{{Ambitwistor Strings in Four
  Dimensions}},
  \href{https://doi.org/10.1103/PhysRevLett.113.081602}{\emph{Phys. Rev. Lett.}
  {\bfseries 113} (2014) 081602}
  [\href{https://arxiv.org/abs/1404.6219}{{\ttfamily 1404.6219}}].

\bibitem{Spradlin:2009qr}
M.~Spradlin and A.~Volovich, \emph{{From Twistor String Theory To Recursion
  Relations}}, \href{https://doi.org/10.1103/PhysRevD.80.085022}{\emph{Phys.
  Rev.} {\bfseries D80} (2009) 085022}
  [\href{https://arxiv.org/abs/0909.0229}{{\ttfamily 0909.0229}}].

\bibitem{Lipstein:2015rxa}
A.~E. Lipstein, \emph{{Soft Theorems from Conformal Field Theory}},
  \href{https://doi.org/10.1007/JHEP06(2015)166}{\emph{JHEP} {\bfseries 06}
  (2015) 166} [\href{https://arxiv.org/abs/1504.01364}{{\ttfamily
  1504.01364}}].

\bibitem{Huang:2015yka}
R.~Huang, J.~Rao, B.~Feng and Y.-H. He, \emph{{An Algebraic Approach to the
  Scattering Equations}},
  \href{https://doi.org/10.1007/JHEP12(2015)056}{\emph{JHEP} {\bfseries 12}
  (2015) 056} [\href{https://arxiv.org/abs/1509.04483}{{\ttfamily
  1509.04483}}].

\bibitem{Sogaard:2015dba}
M.~Søgaard and Y.~Zhang, \emph{{Scattering Equations and Global Duality of
  Residues}}, \href{https://doi.org/10.1103/PhysRevD.93.105009}{\emph{Phys.
  Rev.} {\bfseries D93} (2016) 105009}
  [\href{https://arxiv.org/abs/1509.08897}{{\ttfamily 1509.08897}}].

\bibitem{DeLaurentis:2019phz}
G.~De~Laurentis and D.~Maître, \emph{{Extracting analytical one-loop
  amplitudes from numerical evaluations}},
  \href{https://doi.org/10.1007/JHEP07(2019)123}{\emph{JHEP} {\bfseries 07}
  (2019) 123} [\href{https://arxiv.org/abs/1904.04067}{{\ttfamily
  1904.04067}}].

\bibitem{Farrow:2018cqi}
J.~A. Farrow, \emph{{A Monte Carlo Approach to the 4D Scattering Equations}},
  \href{https://doi.org/10.1007/JHEP08(2018)085}{\emph{JHEP} {\bfseries 08}
  (2018) 085} [\href{https://arxiv.org/abs/1806.02732}{{\ttfamily
  1806.02732}}].

\bibitem{Dolan:2014ega}
L.~Dolan and P.~Goddard, \emph{{The Polynomial Form of the Scattering
  Equations}}, \href{https://doi.org/10.1007/JHEP07(2014)029}{\emph{JHEP}
  {\bfseries 07} (2014) 029} [\href{https://arxiv.org/abs/1402.7374}{{\ttfamily
  1402.7374}}].

\bibitem{Dolan:2015iln}
L.~Dolan and P.~Goddard, \emph{{General Solution of the Scattering Equations}},
  \href{https://doi.org/10.1007/JHEP10(2016)149}{\emph{JHEP} {\bfseries 10}
  (2016) 149} [\href{https://arxiv.org/abs/1511.09441}{{\ttfamily
  1511.09441}}].

\bibitem{Cardona:2015ouc}
C.~Cardona and C.~Kalousios, \emph{{Elimination and recursions in the
  scattering equations}},
  \href{https://doi.org/10.1016/j.physletb.2016.03.003}{\emph{Phys. Lett.}
  {\bfseries B756} (2016) 180}
  [\href{https://arxiv.org/abs/1511.05915}{{\ttfamily 1511.05915}}].

\bibitem{Cachazo:2014xea}
F.~Cachazo, S.~He and E.~Y. Yuan, \emph{{Scattering Equations and Matrices:
  From Einstein To Yang-Mills, DBI and NLSM}},
  \href{https://doi.org/10.1007/JHEP07(2015)149}{\emph{JHEP} {\bfseries 07}
  (2015) 149} [\href{https://arxiv.org/abs/1412.3479}{{\ttfamily 1412.3479}}].

\bibitem{Azevedo:2017lkz}
T.~Azevedo and O.~T. Engelund, \emph{{Ambitwistor formulations of R$^{2}$
  gravity and (DF)$^{2}$ gauge theories}},
  \href{https://doi.org/10.1007/JHEP11(2017)052}{\emph{JHEP} {\bfseries 11}
  (2017) 052} [\href{https://arxiv.org/abs/1707.02192}{{\ttfamily
  1707.02192}}].

\bibitem{DeLaurentis:2016abc}
G.~De~Laurentis, \emph{{The CHY formalism for massless scattering (master
  thesis)}},
  {\emph{\href{https://gdelaurentis.github.io/files/CHYReview.pdf}{Oxford MPhys
  thesis}} (2016) }.

\bibitem{Farrow:2018yqf}
J.~A. Farrow and A.~E. Lipstein, \emph{{New Worldsheet Formulae for Conformal
  Supergravity Amplitudes}},
  \href{https://doi.org/10.1007/JHEP07(2018)074}{\emph{JHEP} {\bfseries 07}
  (2018) 074} [\href{https://arxiv.org/abs/1805.04504}{{\ttfamily
  1805.04504}}].

\bibitem{Berger:2008ag}
C.~F. Berger, Z.~Bern, L.~J. Dixon, F.~Febres~Cordero, D.~Forde, H.~Ita et~al.,
  \emph{{One-Loop Calculations with BlackHat}},
  \href{https://doi.org/10.1016/j.nuclphysbps.2008.09.123}{\emph{Nucl. Phys.
  Proc. Suppl.} {\bfseries 183} (2008) 313}
  [\href{https://arxiv.org/abs/0807.3705}{{\ttfamily 0807.3705}}].

\bibitem{Maitre:2007jq}
D.~Maitre and P.~Mastrolia, \emph{{S@M, a Mathematica Implementation of the
  Spinor-Helicity Formalism}},
  \href{https://doi.org/10.1016/j.cpc.2008.05.002}{\emph{Comput. Phys. Commun.}
  {\bfseries 179} (2008) 501}
  [\href{https://arxiv.org/abs/0710.5559}{{\ttfamily 0710.5559}}].

\bibitem{Berkovits:2004jj}
N.~Berkovits and E.~Witten, \emph{{Conformal supergravity in twistor-string
  theory}}, \href{https://doi.org/10.1088/1126-6708/2004/08/009}{\emph{JHEP}
  {\bfseries 08} (2004) 009}
  [\href{https://arxiv.org/abs/hep-th/0406051}{{\ttfamily hep-th/0406051}}].

\bibitem{Johansson:2018ues}
H.~Johansson, G.~Mogull and F.~Teng, \emph{{Unraveling conformal gravity
  amplitudes}}, \href{https://doi.org/10.1007/JHEP09(2018)080}{\emph{JHEP}
  {\bfseries 09} (2018) 080}
  [\href{https://arxiv.org/abs/1806.05124}{{\ttfamily 1806.05124}}].

\bibitem{Maldacena:2011mk}
J.~Maldacena, \emph{{Einstein Gravity from Conformal Gravity}},
  \href{https://arxiv.org/abs/1105.5632}{{\ttfamily 1105.5632}}.

\bibitem{Anastasiou:2016jix}
G.~Anastasiou and R.~Olea, \emph{{From conformal to Einstein Gravity}},
  \href{https://doi.org/10.1103/PhysRevD.94.086008}{\emph{Phys. Rev.}
  {\bfseries D94} (2016) 086008}
  [\href{https://arxiv.org/abs/1608.07826}{{\ttfamily 1608.07826}}].

\bibitem{Johansson:2017srf}
H.~Johansson and J.~Nohle, \emph{{Conformal Gravity from Gauge Theory}},
  \href{https://arxiv.org/abs/1707.02965}{{\ttfamily 1707.02965}}.

\bibitem{Elvang:2013cua}
H.~Elvang and Y.-t. Huang, \emph{{Scattering Amplitudes}},
  \href{https://arxiv.org/abs/1308.1697}{{\ttfamily 1308.1697}}.

\end{thebibliography}\endgroup
\bibliographystyle{JHEP}

\pagebreak
\begin{appendices}
\section{lips (phase space generator)}\label{sec:appendix1}

In this first appendix we present in more details the \href{https://github.com/GDeLaurentis/lips/}{lips} \verb!Python! package. It is an object-oriented high-precision floating-point phase space generator. It is not the focus of this work, but it is necessary in order to pass a sufficiently precise phase space point to the scattering equations solver function or numerical amplitude object.

The \verb!lips! phase space generator is built on two layers. The lower one, called \verb!Particle!, describes the kinematics of a single particle. Though setters and getters, it provides self-updating numerical tensors for the left and right spinors, four vectors and rank two spinors. This means that if, say, the value of the four momentum is changed, then the values of the spinor attributes are immediately recalculated to reflect the change. We can see the naming conventions in the following code snippet:
\vspace{-3mm}
\begin{minted}[escapeinside=||, mathescape=true, linenos=False, numbersep=5pt, gobble=2, frame=lines, framesep=2mm, breaklines, breakautoindent=false, breakindent=-12.5pt, ]{python}
  >>> oParticle = Particle()
  >>> oParticle.l_sp_u    # left spinor with index up ($\bar{\lambda}^{\dot{\alpha}}$)
  >>> oParticle.r_sp_d    # right spinor with index down ($\lambda_{\alpha}$)
  >>> oParticle.four_mom  # four momentum with index up ($P^\mu$)
  >>> oParticle.r2_sp     # rank two spinor ($P^{\dot{\alpha}\alpha}$)
\end{minted}

\noindent By default the \verb!Particle! object is initialised with random complex momenta. However, this can be overruled by specifying the optional paramenter \verb!real_momentum=True!. A custom value for any of these attributes can also be passed. For instance, we can set the momentum to be along the x axis:
\vspace{-3mm}
\begin{minted}[escapeinside=||, mathescape=true, linenos=False, numbersep=5pt, gobble=2, frame=lines, framesep=2mm, breaklines, breakautoindent=false, breakindent=-12.5pt, ]{python}
  >>> oParticle.four_mom = [1, 1, 0, 0]
\end{minted}

The second layer is a list subclass, called \verb!Particles!. It is a base-one list of Particle objects with several methods attached associated to it. The reason why the list is rebased to start from 1 instead of 0 is simply to match the notation in the amplitudes community. As we have observed, it is initialised as follows:
\vspace{-3mm}
\begin{minted}[escapeinside=||, mathescape=true, linenos=False, numbersep=5pt, gobble=2, frame=lines, framesep=2mm, breaklines, breakautoindent=false, breakindent=-12.5pt, ]{python}
  >>> oParticles = Particles(6)  # argument is the multiplicity
\end{minted}

It also accepts an optional parameter, now called \verb!real_momenta!, which is by default set to \verb!False!, and which gets automatically passed down to all the \verb!Particle! objects in the \verb!Particles! list, thus generating a complex or real phase space point.

\makebox[\textwidth - 20pt][s]{Furthermore, as discussed in conjunction with the analytical reconstruction, the} \linebreak \verb!Particles! phase space can be manipulated to generate specific configurations. For instance, we can generate phase space point with vanishing angle bracket $⟨12⟩$ by calling:
\vspace{-3mm}
\begin{minted}[escapeinside=||, mathescape=true, linenos=False, numbersep=5pt, gobble=2, frame=lines, framesep=2mm, breaklines, breakautoindent=false, breakindent=-12.5pt, ]{python}
  >>> oParticles.set("⟨1|2⟩", 10 ** -30)
\end{minted}

Doubly singular limits for pairs of invariants can be similarly generated. For instance, we can make both $⟨12⟩$ and $⟨23⟩$ small:
\vspace{-3mm}
\begin{minted}[escapeinside=||, mathescape=true, linenos=False, numbersep=5pt, gobble=2, frame=lines, framesep=2mm, breaklines, breakautoindent=false, breakindent=-12.5pt, ]{python}
  >>> oParticles.set_pair("⟨1|2⟩", 10 ** -30, "⟨2|3⟩", 10 ** -30)
\end{minted}

At present these functions only work with complex momenta, because with complex momenta it is possible to construct phase space points where, say, $⟨12⟩$ is small but $[12]$ is not, while with real momenta this is not possible ($[12] \sim ⟨12⟩^*$).

Other notable functions are:
\vspace{-3mm}
\begin{minted}[escapeinside=||, mathescape=true, linenos=False, numbersep=5pt, gobble=2, frame=lines, framesep=2mm, breaklines, breakautoindent=false, breakindent=-12.5pt, ]{python}
  >>> oParticles.randomise_all()  # randomises all momenta
  >>> oParticles.angles_for_squares()  # swaps right/left spinors (C-sym.)
  >>> oParticles.image("234561")  # argument is a permutation of 123...n
\end{minted}

\makebox[\textwidth - 20pt][s]{For more details we refer the reader to the package documentation on the github pages} \linebreak at \href{https://gdelaurentis.github.io/lips/}{lips}.

\pagebreak
\section{seampy (further details)}\label{sec:appendix2}

In this appendix we provide more details on the \href{https://github.com/GDeLaurentis/seampy/}{seampy} package. Although not crucial from a user point of view, these may be of interest if one wants to study in more details the internal behaviour of the program or perform modifications, such as add new theories to the list of those available for computation.

Still using $n=6$ for our examples, we can see two important elements of the elimination theory algorithm:

$\circ$ the vector of variables to be removed via elimination theory from Eq.~(\ref{eq:extended-V}):
\vspace{-3mm}
\begin{minted}[escapeinside=||, mathescape=true, linenos=False, numbersep=5pt, gobble=2, frame=lines, framesep=2mm, breaklines, breakautoindent=false, breakindent=-12.5pt, ]{python}
  >>> V(6)
  [1, z₂, z₃, z₂⋅z₃, z₃², z₂⋅z₃²]
\end{minted}

$\circ$ the elimination theory matrix obtained with the recursion algorithm of Eq.~(\ref{eq:matrix-recursion}):
\vspace{-3mm}
\begin{minted}[escapeinside=||, mathescape=true, linenos=False, numbersep=5pt, gobble=2, frame=topline, framesep=2mm, breaklines, breakautoindent=false, breakindent=-12.5pt, ]{python}
  >>> M(6)
\end{minted}
\vspace{-7mm}
\begin{center}
  \noindent\includegraphics{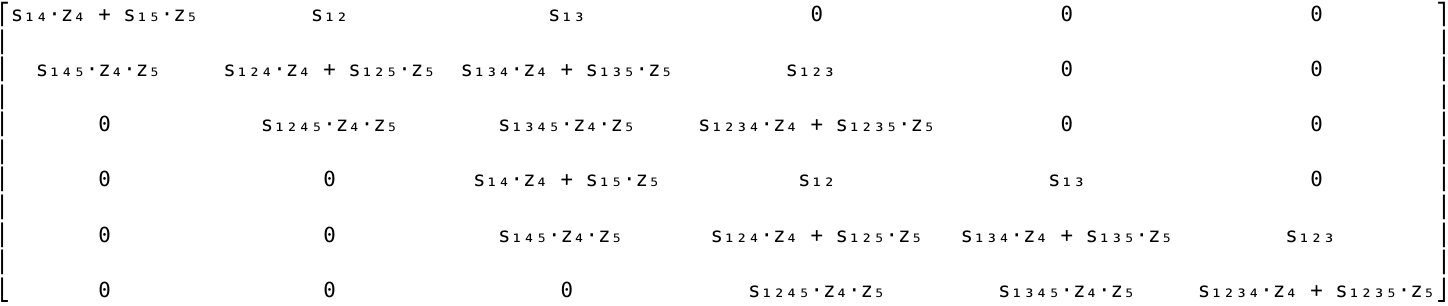}
\end{center}
\vspace{-14mm}
\begin{minted}[escapeinside=||, mathescape=true, linenos=False, numbersep=5pt, gobble=2, frame=bottomline, framesep=2mm, breaklines, breakautoindent=false, breakindent=-12.5pt, ]{python}
\end{minted}


These are the basis for the \verb!solve_scattering_equations! function, which involves taking the determinant of \verb!M! and finding its roots.

We can also consider the CHY-integrands and the Jacobian for the change of variables. We denote with the term \textit{reduced} the following sequence of operations: a) removing rows and columns: two of them for arguments of Pfaffians and three of them for the Jacobian; b) imposing the M\"obius fixing choice of Eq.~(\ref{eq:mobius-fix-choice}); c) removing any factorised factor of $z_1 = \infty$. In the following code snippets we reproduce some examples:

$\circ$ the reduced Jacobian Matrix $\phi$ of Eq.~(\ref{eq:jacobian-matrix}):
\vspace{-3mm}
\begin{minted}[escapeinside=||, mathescape=true, linenos=False, numbersep=5pt, gobble=2, frame=topline, framesep=2mm, breaklines, breakautoindent=false, breakindent=-12.5pt, ]{python}
  >>> Phi(6)
\end{minted}
\vspace{-7mm}
\begin{center}
  \noindent\includegraphics{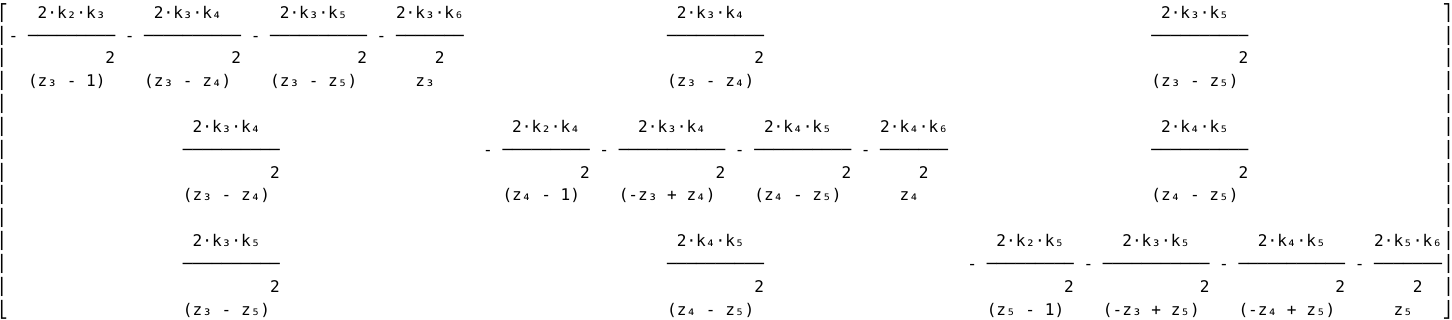}
\end{center}
\vspace{-14mm}
\begin{minted}[escapeinside=||, mathescape=true, linenos=False, numbersep=5pt, gobble=2, frame=bottomline, framesep=2mm, breaklines, breakautoindent=false, breakindent=-12.5pt, ]{python}
\end{minted}

\pagebreak

\indent $\circ$ the reduced matrix A of Eq.~(\ref{eq:Psi-A}):
\vspace{-3mm}
\begin{minted}[escapeinside=||, mathescape=true, linenos=False, numbersep=5pt, gobble=2, frame=topline, framesep=2mm, breaklines, breakautoindent=false, breakindent=-12.5pt, ]{python}
  >>> A(6)
\end{minted}
\vspace{-3mm}
\noindent\hspace*{0.8em}\includegraphics{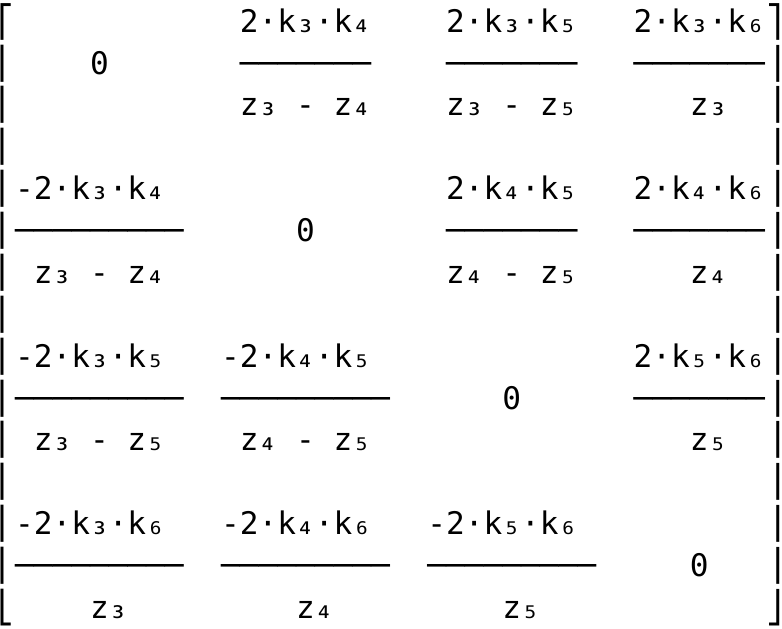}
\vspace{-9mm}
\begin{minted}[escapeinside=||, mathescape=true, linenos=False, numbersep=5pt, gobble=2, frame=bottomline, framesep=2mm, breaklines, breakautoindent=false, breakindent=-12.5pt, ]{python}
\end{minted}


$\circ$ the reduced cyclic Parke-Taylor-like factor $C_n$ of Eq.~(\ref{eq:cyclic-factor}):
\vspace{-3mm}
\begin{minted}[escapeinside=||, mathescape=true, linenos=False, numbersep=5pt, gobble=2, frame=lines, framesep=2mm, breaklines, breakautoindent=false, breakindent=-12.5pt, ]{python}
  >>> Cyc(6)
                 -1                
  ────────────────────────────────
  z₅⋅(-z₃ + 1)⋅(z₃ - z₄)⋅(z₄ - z₅)
\end{minted}

All these symbolic quantities are built with \href{https://www.sympy.org/en/index.html}{sympy}. However, note that the symbolical substitution function from sympy is very slow, therefore we use regular expressions from the \href{https://docs.python.org/2/library/re.html}{re} library to perform substitutions in the conversion from symbolic to numeric.

For more details we refer the reader to the package documentation on the github pages at \href{https://gdelaurentis.github.io/seampy/}{seampy}.

\end{appendices}

\end{document}